\documentclass[a4paper,12pt]{article}
\pdfoutput=1 

\usepackage{jheppub} 

\usepackage[T1]{fontenc} 
\usepackage{multirow}
\usepackage{tikz}

\usepackage{pdflscape}

\usetikzlibrary{matrix, positioning}

\usepackage{slashed}
\usepackage{subcaption}

\def\ep{\epsilon}
\def\varep{\varepsilon}
\def\mt{m_{t}}

\def\ubar{\overline{u}}

\def\kt{{p_4}}
\def\skt{\slashed{p}_4}
\def\kb{{p_2}}
\def\ku{{p_1}}
\def\sku{\slashed{p}_1}
\def\kd{{p_3}}

\def\S#1{{\cal S}_{#1}}

\def\vS{\vec{\cal{S}}}
\def\as{\alpha_{s}}

\def\bra#1{\langle #1 \vert}
\def\ket#1{\vert #1 \rangle}
\def\braket#1#2{\langle #1 \vert #2 \rangle}

\def\as{\alpha_s}

\def\anodim{\boldsymbol{\Gamma}}

\newcommand{\be}{\begin{equation}}
\newcommand{\ee}{\end{equation}}

\title{ On non-factorisable contributions  to  $t$-channel single-top production}

\author{Christian Br\o{}nnum-Hansen,}
\author{Kirill Melnikov,}
\author{J\'{e}r\'{e}mie Quarroz,}
\author{Chen-Yu Wang}

\affiliation{Institute for Theoretical Particle Physics, KIT, Karlsruhe, Germany}

\emailAdd{christian.broennum-hansen@kit.edu}
\emailAdd{kirill.melnikov@kit.edu}
\emailAdd{jeremie.quarroz@kit.edu}
\emailAdd{chen-yu.wang@kit.edu}

\abstract{We compute  the non-factorisable contribution
  to the two-loop helicity amplitude for  $t$-channel single-top production, the last missing
  piece of the two-loop virtual corrections to  this process. 
  Our calculation employs  analytic reduction to master integrals 
  and  the auxiliary mass flow method for their fast  numerical evaluation. We study the   impact
  of these corrections on basic observables that are measured
experimentally in  the single-top production process.}

\preprint{TTP21-027, P3H-21-057}

\begin{document}
\maketitle
\flushbottom

\section{Introduction}

Hadronic production of top quarks at the LHC provides an opportunity to study the  heaviest particle
of the Standard Model in great detail. Since, according to the Standard Model, top quarks receive
their masses exclusively through interactions with the Higgs background field, a better understanding of top quark
properties may lead to  a better understanding of electroweak symmetry breaking in and, hopefully,
beyond, the Standard Model. 

At  a hadron collider top quarks and anti-quarks are primarily produced in
pairs by means of strong interactions. However, single-top production, which necessarily involves the
weak $tWb$ interaction vertex,   also occurs  quite frequently at the LHC. In fact,
the single-top production cross section at the LHC is  about a quarter of the cross section
to produce a $t \bar t$ pair. Such a large cross section and an impressive 
luminosity  collected at the LHC  implies that   by now ${\cal O}(10)$  millions top quarks have been produced
there  thanks to this mechanism.

The interest in single-top production is related to the fact that
weak interactions are responsible for this process.  This opens up  a number 
of interesting opportunities~\cite{Giammanco:2017xyn}
that involve studies of the structure of the $tbW$ vertex \cite{ATLAS:2017ygi,ATLAS:2019hhu},
improving constrains on the CKM matrix elements \cite{CMS:2020vac,ATLAS:2019hhu} and indirect  determination  of the top quark
width $\Gamma_t$ \cite{CMS:2014mxl}.  More recently, measurements of the top quark mass in single-top events
started  to play a more visible role in  the top quark mass measurements at the LHC \cite{CMS:2017mpr}. Finally, detailed
studies of QCD dynamics in single-top  production processes including interesting constraints
on parton distribution functions and precise measurements of kinematic distributions are benefitting
from the high integrated luminosity of the LHC \cite{CMS:2019jjp,ATLAS:2017rso}. 

At a hadron collider, single top quarks  can be produced in three different ways (for a review, see Ref.~\cite{Giammanco:2017xyn}).  One distinguishes 
{\it i})  the  $t$-channel process that refers to 
$q\, b \to q'\, t$ scattering  mediated by an exchange of a $W$ boson,
{\it ii})  the  $s$-channel process that at the partonic level
corresponds to $q \, \overline{q'} \to W^*  \to t\, \overline{b}$
and, finally,  {\it iii}) the associated production that involves the $g\, b \to W\, t$ process. 
About $70 \%$ of single top quarks at the LHC  are  produced in the 
$t$-channel process; ${\cal O}(25 \%)$ are due to  the associated $tW$ 
production and only ${\cal O}(5\%)$ are due to the $s$-channel process.

Studies  of  single-top production  rely on a precise  theoretical description of this process  that can be
obtained in the context of perturbative QCD and collinear factorisation.
This has been done at 
next-to-leading order (NLO) in perturbative
QCD in Refs.~\cite{Harris:2002md,Campbell:2004ch,Sullivan:2004ie,Cao:2004ky,Schwienhorst:2010je,Gao:2021plf}.
Furthermore, for the $t$-channel production 
next-to-next-to-leading order (NNLO) QCD corrections  have been calculated  in
Refs.~\cite{Brucherseifer:2014ama,Berger:2016oht,Campbell:2020fhf}. Although the more recent computations 
of such corrections presented in Refs.~\cite{Berger:2016oht,Campbell:2020fhf}  are quite sophisticated
and incorporate top quark decays and QCD corrections to them in the narrow width approximation, 
all existing calculations of  NNLO QCD corrections to  $t$-channel single-top production 
do not account for the so-called  {\it non-factorisable}  contributions.

\begin{figure}
	\centering
	\includegraphics[scale=0.6]{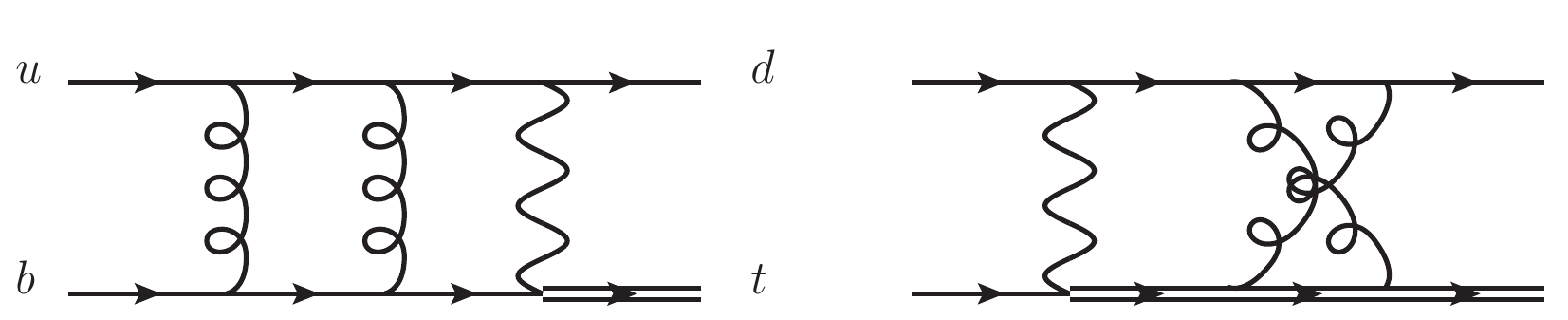}
	\caption{Examples of  non-factorisable two-loop diagrams.
          Wavy lines represent $W$ bosons, curly lines are gluons, solid lines are quarks. The double line represents the top quark.}\label{fig:diagrams}
\end{figure}

In the context of $t$-channel single-top  production,  non-factorisable corrections refer
to contributions that connect a light-quark line and a heavy $b \to t$ line by gluon exchanges, see Figure~\ref{fig:diagrams}.
Thanks to colour conservation, such contributions vanish when 
NLO QCD predictions for cross sections are computed.
However, since at next-to-next-to-leading order two gluons in a colour-singlet state can be exchanged
between different fermion lines,  non-factorisable
diagrams start contributing at that order and, in principle,  have to be accounted for. 

\begin{figure}
	\centering
	\includegraphics[scale=0.6]{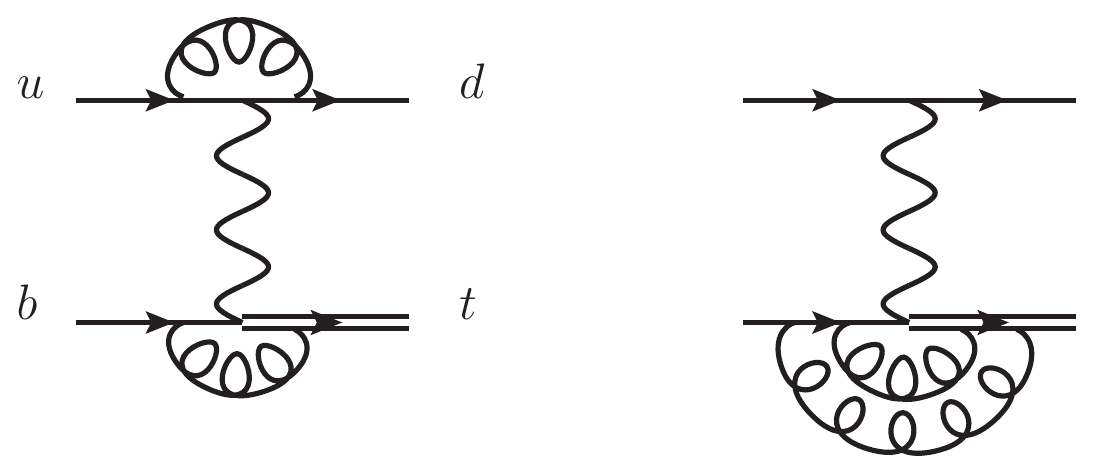}
	\caption{Examples of  factorisable two-loop diagrams not considered in the present calculation.
          Wavy lines represent $W$ bosons, curly lines are gluons, solid lines are quarks. The double line represents the top quark.}\label{fig:factorisable}
\end{figure}

 However, it is far from  obvious  that    these non-factorisable corrections  are important
 for a precise description  of   single-top production. The reason for neglecting them in earlier
 computations was that  they are colour-suppressed compared to factorisable contributions shown in Figure~\ref{fig:factorisable}.
 On the other hand, as became clear recently, these non-factorisable corrections  may be enhanced
 by a  factor  $\pi^2$  related to remnants of the so-called
 Coulomb or Glauber phase \cite{glauber}. Indeed, the existence of such
 an enhancement  was recently demonstrated \cite{Liu:2019tuy} in the context of Higgs boson production in weak boson
 fusion.  In fact it was shown in that reference that the $\pi^2$-enhancement of non-factorisable
 corrections largely compensates  their ${\cal O}(1/N_c^2)$
 suppression, so that the non-factorisable corrections to  Higgs production in weak boson fusion are 
 larger than the  colour-suppression argument suggests.  Moreover, it is known that
 factorisable NNLO QCD corrections to single-top production cross section and basic kinematic 
 distributions  are rather  small~\cite{Brucherseifer:2014ama,Berger:2016oht,Campbell:2020fhf}. This 
 smallness of factorisable QCD corrections makes non-factorisable corrections more relevant provided,
 of course, that high-precision  theoretical description of single-top production is of interest. 

 The goal of this paper is to make the first step towards a better understanding of  non-factorisable
 corrections to single-top production at the LHC and to calculate their contributions
 to  the two-loop virtual amplitude.  We do this by expressing all  two-loop integrals that appear in
 non-factorisable diagrams 
through  master integrals keeping
 exact dependence on the top quark mass and the $W$ mass 
 and by computing these integrals using the auxiliary mass flow  method~\cite{Liu:2017jxz,Liu:2020kpc,Liu:2021wks}.\footnote{We note that
   the very first reduction of the non-factorisable contributions to single-top production 
   to master integrals was performed in Ref.~\cite{Assadsolimani:2014oga}, albeit
   for a   fixed numerical relation between the top-quark mass and the $W$ boson mass
   $m_t^2 = 14m_W^2/3$.  Furthermore, a reduction of planar, non-factorisable diagrams for $W$-associated
   single-top production was recently presented  in Ref.~\cite{Basat:2021xnn}.}
 As we explain in detail below, this computational set up is similar to the one  used  previously by two of the
 present authors~\cite{Bronnum-Hansen:2020mzk,Bronnum-Hansen:2021olh}.

 This paper is organised as follows. In Section~\ref{sec:ampcalc} we discuss technical details
 pertinent to  the calculation of non-factorisable contributions to the single-top production amplitude. 
 In Section~\ref{sec:integral} we describe the numerical evaluation of the master  integrals.
 The (infrared) pole structure of the non-factorisable contribution to the amplitude is discussed in Section~\ref{sec:polestruct}.
 The impact of non-factorisable corrections 
 on  the cross section and some kinematic distributions
 are studied  in Section~\ref{sec:results}.  We conclude in Section~\ref{sec:conclusions}.
 Numerical values for non-factorisable contributions to the two-loop amplitude at a few kinematic points
 are presented in Appendix~\ref{app:evaluations}.
 Boundary conditions for master integrals that we used in this calculation can be found in an ancillary file.

\section{Non-factorisable contributions to helicity amplitudes }\label{sec:ampcalc}

We consider single-top production in the $t$-channel and, for definiteness, focus on a particular flavour of light quarks 
\begin{align}
  u(p_1) + b(p_2) &\to d(p_3) + t(p_4).
  \label{eq:process}
\end{align}
Except for the top quark, all other quarks in Eq.~\eqref{eq:process} are massless, so that  $p_i^2 = 0,\ i=1,2,3$. The top quark
is on the mass-shell  $p_4^2 = m_t^2$.
We follow standard conventions and define Mandelstam variables as 
\begin{align}
s = (p_1 + p_2)^2,\qquad
t = (p_1 - p_3)^2,\qquad
u = (p_2 - p_3)^2,
\end{align}
with $ s+ t+u = m_t^2$.

We write the  amplitude of the process in Eq.~\eqref{eq:process} expanded
in the {\it renormalised} strong coupling constant  $\as = \as(\mu)$ as follows
\begin{align}
  \mathcal{A}(\{ p_i \}) =
 g_w^2 V_{ud} V_{tb} \left ( 
    \mathcal{A}^{(0)} + \frac{\as}{4\pi}
     \mathcal{A}_{\rm nf}^{(1)} + \left(\frac{\as}{4\pi}\right)^2
     \mathcal{A}_{\rm nf}^{(2)}+... + \mathcal{O}\left(\as^3\right)
     \right )
    \,.\label{eq:asexp}
\end{align}
When writing Eq.~\eqref{eq:asexp},  we have extracted
the weak coupling constant $g_w$
and the CKM matrix elements
$V_{tb}$ and $V_{ud}$.  Also,
 $\mathcal{A}^{(0)} = \mathcal{A}^{(0)}(\{ p_i \})$
is the (properly normalised) Born amplitude of the process Eq.~\eqref{eq:process}, 
$\mathcal{A}_{\rm nf}^{(1,2)} = \mathcal{A}_{\rm nf}^{(1,2)}(\{ p_i \})$ are one- and two-loop non-factorisable amplitudes respectively,
and ellipses stand for factorisable contributions that we do not discuss in this paper.\footnote{
   We note that two-loop factorisable contributions  to the full amplitude
are of the vertex type, see Figure~\ref{fig:factorisable}. For the $udW$ vertex they were computed in Refs.~\cite{Gonsalves:1983nq,vanNeerven:1985xr}, 
whereas for  the  $tb W$ vertex  they were calculated in  Refs.~\cite{Bonciani:2008wf,Bell:2008ws,
  Asatrian:2008uk,Beneke:2008ei,Huber:2009se}.}

\begin{figure}
	\centering
	\includegraphics[scale=0.6]{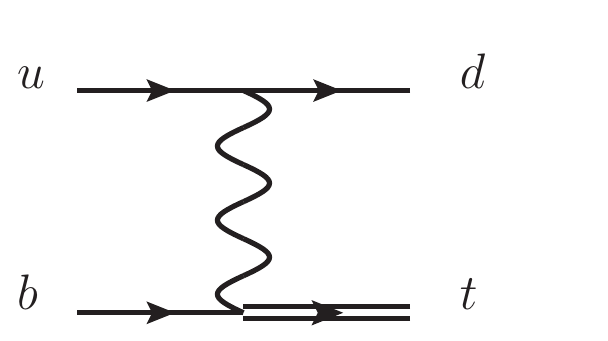}
	\caption{Tree-level diagram. Wavy lines represent $W$ bosons, solid lines are quarks. The double line represents the top quark.}\label{fig:treelevel}
\end{figure}

To proceed further, we perform the colour decomposition of relevant amplitudes.
Figure~\ref{fig:treelevel} shows the only diagram that contributes to $t$-channel single-top  production at tree level.
Since $W$ bosons carry no colour charge, we find  
\be
\mathcal{A}^{(0)} = \hat 1_{c_3 c_1} \hat 1_{c_4 c_2} A^{(0)} = \delta_{c_1 c_3} \delta_{c_2 c_4}\,  A^{(0)}, 
\ee
where $\hat 1$ is the identity  matrix, 
$c_{1,..,4}$ are the colour indices of particles with momenta $p_{1,..,4}$,  respectively, and $A^{(0)}$ is the colour-stripped amplitude.

Four box diagrams with identical colour factors contribute to the one-loop non-factorisable amplitude. We write 
\begin{align}
  \mathcal{A}_{\rm nf}^{(1)} = T^a_{c_3 c_1} T^a_{c_4 c_2} A_{\rm nf}^{(1)} = \frac{1}{2}
  \left( \delta_{c_1 c_4} \delta_{c_2 c_3} - \frac{1}{N_c}  \delta_{c_1 c_3} \delta_{c_2 c_4}   \right) A_{\rm nf}^{(1)}.
\end{align}
We note that  the interference of the one-loop amplitude and the Born amplitude vanishes thanks to colour conservation
\begin{align}
\sum_{\text{colour}} \mathcal{A}^{(0) \star} \mathcal{A}_{\rm nf}^{(1)} = 0.
\end{align}

At two loops  eighteen  non-factorisable  box diagrams need to be considered; we generate them
using \texttt{QGRAF}~\cite{Nogueira:1991ex}. Since $W$ bosons are colourless, these diagrams are
of both planar and non-planar types as far as QCD interactions are concerned.  For this reason, there are just two
distinct colour factors
\be
c_{2,\rm pl} = (T^a T^b )_{c_3 c_1} (T^a T^b)_{c_4 c_2},\;\;\;\; c_{2,\rm npl} = (T^a T^b )_{c_3 c_1} (T^b T^a)_{c_4 c_2},\;\;\;\;
\ee
so that
\be
\mathcal{A}_{\rm nf}^{(2)} =
c_{2,\rm pl} A_{\rm nf}^{(2),\rm pl} + c_{2,\rm npl} A_{\rm nf}^{(2),\rm npl}.
\ee
The two amplitudes $A_{\rm nf}^{(2),\rm pl}$ and $A_{\rm nf}^{(2),\rm npl}$ are  obtained by computing (QCD) planar and non-planar diagrams, respectively.

However, it is easy to realise that only a particular combination of these amplitudes
contributes to  NNLO QCD
cross section through interference  with the leading-order amplitude. Indeed, since the leading-order colour factor
involves $\delta_{c_3 c_1} \delta_{c_4 c_2}$, when the interference of non-factorisable two-loop diagrams and
the tree amplitude is computed,
we obtain ${\rm Tr}(T^{a_1} T^{a_2})$ for each of the fermion lines. However, since   ${\rm Tr}(T^a T^b) = {\rm Tr}(T^b T^a)$,
the distinction between colour factors for planar and non-planar diagrams disappears.  To project on the relevant structure,
we write
\be
2 T^a T^b = \{T^a,T^b\} + [T^a,T^b],\;\;\; 2 T^b T^a = \{T^a,T^b\} - [T^a,T^b].\;\;\; 
\ee
Since ${\rm Tr}\left( [T^a,T^b] \right)=0$, commutators of colour generators do not contribute to the interference.
As the
result,  we can write 
\be
\begin{split} 
\mathcal{A}_{\rm nf}^{(2)} & = \frac{1}{4}
\{T^a,T^b\}_{c_3 c_1}   \{T^a,T^b\}_{c_4 c_2} \;  ( A_{\rm nf}^{(2),\rm pl} +  A_{\rm nf}^{2,\rm npl} ) + \dots
\\
&  = \frac{1}{4} \{T^a,T^b\}_{c_3 c_1}   \{T^a,T^b\}_{c_4 c_2} \; A_{\rm nf}^{(2)} + \dots,
\label{eq4.10a}
\end{split} 
\ee
where ellipses stand for terms that vanish when the interference of $\mathcal{A}_{\rm nf}^{(2)}$ with tree amplitude
is computed. We note that  we  introduced  $A_{\rm nf}^{(2)}  = A_{\rm nf}^{(2),\rm pl} +  A_{\rm nf}^{2,\rm npl}$ in Eq.~\eqref{eq4.10a}.
We find
\be
  \sum_{\text{colour}}   \mathcal{A}^{(0)*} \mathcal{A}_{\rm nf}^{(2)}
  = \frac{1}{4} (N_c^2 - 1) \; A^{(0)*} A_{\rm nf}^{(2)}. \label{eq:interference}
\ee

To compute  relevant one- and two-loop amplitudes,  we need to write them in terms of invariant form factors
and independent Lorentz structures.    Since charged weak currents involve left-handed
projectors and, therefore, the Dirac matrix $\gamma_5$, care is needed when performing computations
in dimensional regularisation. However, since no closed fermion loops contribute
to non-factorisable corrections, we can make use of an anti-commuting prescription for the $\gamma_5$
and move left-handed projectors to act on  the external {\it massless} fermion states.
It then becomes clear that we can consider amplitudes mediated by the \emph{vector} current
but  only account for left-handed massless quarks when constructing physical 
amplitudes for the charged current.

\allowdisplaybreaks

There are   eleven structures that may contribute to the non-factorisable part
of the amplitude through NNLO QCD. They  are\footnote{We use slightly different tensor structures
 as compared to the ones used in Ref.~\cite{Assadsolimani:2014oga}.}
\begin{align}
\begin{split}
  &\S{1} = \ubar_t(\kt)\,\,u(\kb)\times
  \ubar(\kd)\, \skt \,u(\ku)\,, \\
  &\S{2} = \ubar_t(\kt)\, \sku  \,u(\kb)\times
  \ubar(\kd)\, \skt  \,u(\ku)\,, \\
  &\S{3} =\ubar_t(\kt)\, \gamma^{\mu_1}  \,u(\kb)\times
  \ubar(\kd)\, \gamma_{\mu_1} \,u(\ku)\,, \\
  &\S{4} = \ubar_t(\kt)\, \gamma^{\mu_{1}} \sku  \,u(\kb)\times
  \ubar(\kd)\, \gamma_{\mu_{1}}  \,u(\ku)\,, \\
  &\S5=\ubar_t(\kt)\, \gamma^{\mu_{1}}\gamma^{\mu_{2}}  \,u(\kb)\times
  \ubar(\kd)\, \gamma_{\mu_{1}}\gamma_{\mu_{2}} \skt  \,u(\ku)\,, \\
  &\S6=\ubar_t(\kt)\, \gamma^{\mu_{1}}\gamma^{\mu_{2}} \sku  \,u(\kb)\times
   \ubar(\kd)\, \gamma_{\mu_{1}}\gamma_{\mu_{2}} \skt  \,u(\ku)\,,
   \\
  &\S7=\ubar_t(\kt)\, \gamma^{\mu_{1}}\gamma^{\mu_{2}}\gamma^{\mu_{3}}  \,u(\kb)\times
  \ubar(\kd)\, \gamma_{\mu_{1}}\gamma_{\mu_{2}}\gamma_{\mu_{3}} \,u(\ku)\,, \\
  &\S8=\ubar_t(\kt)\, \gamma^{\mu_{1}}\gamma^{\mu_{2}}\gamma^{\mu_{3}} \sku  \,u(\kb)\times
  \ubar(\kd)\, \gamma_{\mu_{1}}\gamma_{\mu_{2}}\gamma_{\mu_{3}}  \,u(\ku)\,, \\
  &\S{9}=\ubar_t(\kt)\, \gamma^{\mu_{1}} \gamma^{\mu_{2}}\gamma^{\mu_{3}}\gamma^{\mu_{4}}  \,u(\kb)\times
  \ubar(\kd)\, \gamma_{\mu_{1}}\gamma_{\mu_{2}}\gamma_{\mu_{3}}\gamma_{\mu_{4}} \skt  \,u(\ku)\,, \\
  &\S{10}=\ubar_t(\kt)\, \gamma^{\mu_{1}}\gamma^{\mu_{2}}\gamma^{\mu_{3}}\gamma^{\mu_{4}} \sku   \,u(\kb)\times
  \ubar(\kd)\, \gamma_{\mu_{1}}\gamma_{\mu_{2}}\gamma_{\mu_{3}}\gamma_{\mu_{4}} \skt  \,u(\ku)\,, \\
  &\S{11}=\ubar_t(\kt)\, \gamma^{\mu_{1}}\gamma^{\mu_{2}}\gamma^{\mu_{3}}\gamma^{\mu_{4}} \gamma^{\mu_{5}}  \,u(\kb)\times
  \ubar(\kd)\, \gamma_{\mu_{1}}\gamma_{\mu_{2}}\gamma_{\mu_{3}}\gamma_{\mu_{4}}  \gamma_{\mu_{5}}   \,u(\ku)\,,  \label{eq:tensorstructs}
  \end{split}
 \end{align} 
where $u_t(\kt)$ denotes the  only  {\it massive} spinor. We note that the above quantities depend on the polarisation
states of external fermions that, in what follows,
we will  denote by $\vec \lambda$. Therefore, we will write  $\S{i} = \S{i}(\vec \lambda)$, $i=1, \dots , 11$.

It is clear that not all eleven structures contribute at leading and next-to-leading order
in the perturbative expansion of the amplitude ${\cal A}$. Indeed, 
at  tree level each fermion line has exactly  one Dirac matrix.
As the result,  the colour-stripped tree-level amplitude for the $u+b \to d + t$ process 
can be written as 
\begin{align}
A^{(0)}(\vec \lambda)  = \frac{\S3(\vec \lambda)}{4(t - m_W^2)} \,.
\end{align}
Upon squaring $\mathcal{A}^{(0)}(\vec \lambda) $ and summing  over colours and appropriate polarisation
states of external fermions, we find 
\begin{align}
    \sum_{\vec \lambda ,\rm colours}|\mathcal{A}^{(0)}(\vec \lambda)|^2 =  N_c^2 \frac{4s(s-m_t^2)}{(t-m_W^2)^2}\,.
\end{align}

The one-loop diagrams have at most three $\gamma$-matrices on each fermion line and can therefore be decomposed in terms of the first seven tensor structures.
At two loops we need all eleven structures to express the amplitude
in terms of invariant form factors. We write
\begin{align}
A_{\rm nf}^{(2)} = \vec{f} \cdot  \vec {\mathcal{S}},\label{eq:tensordecomposition}
\end{align}
where we introduced vectors $\vS$ and $\vec f$  to accommodate  eleven
tensor structures $\vS^T = (\S{1},\S{2},\dots,\S{11})$ and eleven form factors, respectively. 

To compute  the form factors, we calculate  eleven quantities 
\begin{equation}
Q_{i}=\sum_{\vec \lambda} \S{i}^{\dagger}(\vec \lambda)\, A^{(2)}_{\rm nf}(\vec \lambda),\;\;\; i = 1,\dots,  11,
\label{eq:projections2l}
\end{equation}
where the sum runs over all  polarisation states of external fermions. We stress that 
since form factors do not depend on   helicities of  external quarks, we do not need to restrict polarisation
states to left-handed ones when computing the sum in Eq.~\eqref{eq:projections2l}.
Hence, we can use simple formulas to describe density  matrices of external quarks
\begin{align}
  \sum_{\lambda} u (p_i) \otimes \overline{u} (p_i) = \slashed{p}_i,\;\;\;\; i = 1,2,3,\;\;\;\qquad
  \sum_{\lambda} u_t (p_4) \otimes \overline{u}_t (p_4)= \slashed{p}_4 + m_t\,.
\end{align}
For each Feynman diagram that contributes to $A_{\rm nf}^{(2)}$ polarisation sums produce  independent
traces for the two fermion lines. Once these traces are computed,  the results depend on scalar
products of the loop momenta and external momenta and no external spinors are   present 
anymore. At this point, one can define families of integrals and use integration-by-parts identities to
express all the relevant integrals through a relatively small set of master integrals. We describe this point in  detail
in the next section. 
 
To relate the quantities $Q_{i}$ to form factors, we use the representation of the amplitude
in terms of form factors and write 
\be
Q_{i} = \sum_{\vec \lambda} \S{i}^{\dagger}(\vec \lambda)\, A_{\rm nf}^{(2)}(\vec \lambda)
= \sum \limits_{j} f_j \sum_{\vec \lambda} \S{i}^{\dagger}(\vec \lambda)\, \S{j}(\vec \lambda)
= \sum \limits_{j} C_{ij} f_j\,,
\label{eq:projection}
\ee
where the coefficients  $C_{ij}$ read
\be
C_{ij} = \sum_{\vec \lambda}  \S{i}^{\dagger}(\vec \lambda)\, \S{j}(\vec \lambda)\,.
\ee
Turning to vector notation, we rewrite Eq.~\eqref{eq:projection} as 
\be
\vec Q = \hat C \; \vec f\,.
\ee
It follows that 
\be
\vec f = {\hat C}^{-1} \; \vec Q\,.
\ee
This equation allows us to compute the form factors as linear combinations of the amplitude projections $Q_{i}$.

It remains to explain how helicity amplitudes are computed.  To  this end, we make use of the fact that
the four-momenta $p_{1,2,3,4}$ are four-dimensional. This allows us to define polarisation  states of the external
fermions in the standard way. However, since  the Lorentz indices that appear in Eq.~\eqref{eq:tensorstructs}
are $d$-dimensional,  before we can calculate helicity amplitudes we need to remove all Dirac matrices with $(d-4)$-dimensional
indices from these expressions. This can be done if one notices that, to be non-vanishing,  a matrix element between two
``four-dimensional'' spinors requires an {\it even} number of matrices with $(d-4)$-dimensional indices.
This observation allows us to decompose the original tensor structures in terms of their ``four-dimensional'' counter-parts. We find 
\be
\begin{split} 
  \S{1,..,4} &= \S{1,..,4}^{(4)}\,,
  \\
  \S{5,6} &= \S{5,6}^{(4)} - 2\ep \S{1,2}^{(4)}\,,
  \\
  \S{7,8} &= \S{7,8}^{(4)} -6\ep  \S{3,4}^{(4)}\,,
\\
  \S{9,10} &= \S{9,10}^{(4)} -12 \ep  \S{5,6}^{(4)} + \left(12 \ep^2  + 4\ep  \right) \S{1,2}^{(4)}\,,
  \\
  \S{11} &= \S{11}^{(4)} -20 \ep  \S7^{(4)} + \left( 60 \ep^2 + 20\ep  \right) \S3^{(4)}\,,
  \label{eq:tensorstructs4d}
\end{split} 
\ee
where the notation $\S{1,\dots,11}^{(4)}$  refers to the structures shown in Eq.~\eqref{eq:tensorstructs} with all dummy indices 
restricted to four dimensions. Thanks to this restriction, computing
helicity amplitudes using Lorentz structures that appear on the right-hand side
of Eq.~\eqref{eq:tensorstructs4d} is straightforward and unambiguous.

\section{Master integrals}
\label{sec:integral}

To compute the eleven quantities $Q_{i}$, we classify all contributing integrals into 
integral families   using \texttt{REDUZE 2}~\cite{vonManteuffel:2012np}.
We find that we need  to introduce  18 integral families but  half of them are crossings of the other half.
The integral families can be found in Table~\ref{tab:families}.
The integral reduction is performed analytically using \texttt{KIRA}~\cite{Klappert:2020nbg}.
The computational expense is rather
modest and the most complicated reduction takes about four  days on  $20$ cores.
We find that  428 master integrals are required to compute the non-factorisable corrections to single-top production at two loops. 

\begin{table}[ht]
    \centering
    \begin{tabular}{|c|c|p{0.7 \linewidth}|}
        \hline \hline
        \multicolumn{2}{|c|}{\textbf{Name}} & \multicolumn{1}{|c|}{\textbf{Definition}} \\
        \hline \hline
        \multirow{6}{*}{planar}
        & \multirow{2}{*}{1} &
        $
            l_{1}^{2},
            (l_{1} - p_{1})^{2},
            (l_{1} + p_{2})^{2},
            (l_{2} + p_{3})^{2},
            (l_{1} + l_{2} - p_{1} + p_{3})^{2},
        $
        \newline
        $
            (l_{2} - p_{1} - p_{2} + p_{3})^{2},
            l_{2}^{2} - m_{W}^{2},
            l_{1} \cdot p_{3},
            l_{2} \cdot p_{2}
        $
        \\
        \cline{2-3}
        & \multirow{2}{*}{2} &
        $
            l_{1}^{2},
            l_{2}^{2},
            (l_{1} - p_{1})^{2},
            (l_{1} + p_{2})^{2},
            (l_{2} + p_{3})^{2},
        $
        \newline
        $
            (l_{2} - p_{1} - p_{2} + p_{3})^{2} - m_{t}^{2},
            (l_{1} + l_{2} - p_{1} + p_{3})^{2} - m_{W}^{2},
            l_{1} \cdot p_{3},
            l_{2} \cdot p_{2}
        $
        \\
        \cline{2-3}
        & \multirow{2}{*}{3} &
        $
            l_{2}^{2},
            (l_{1} - p_{1})^{2},
            (l_{2} + p_{3})^{2},
            (l_{1} + l_{2} - p_{1} + p_{3})^{2},
            (l_{1} + p_{2})^{2} - m_{t}^{2},
        $
        \newline
        $
            (l_{2} - p_{1} - p_{2} + p_{3})^{2} - m_{t}^{2},
            l_{1}^{2} - m_{W}^{2},
            l_{1} \cdot p_{3},
            l_{2} \cdot p_{2}
        $
        \\
        \hline \hline
        \multirow{12}{*}{non-planar}
        & \multirow{2}{*}{1} &
        $
            l_{2}^{2},
            (l_{2} - p_{1})^{2},
            (l_{1} + p_{3})^{2},
            (l_{1} - l_{2} + p_{3})^{2},
            (l_{1} - l_{2} - p_{2} + p_{3})^{2},
        $
        \newline
        $
            (l_{1} - p_{1} - p_{2} + p_{3})^{2},
            l_{1}^{2} - m_{W}^{2},
            l_{2} \cdot p_{2},
            l_{2} \cdot p_{3}
        $
        \\
        \cline{2-3}
        & \multirow{2}{*}{2} &
        $
            l_{1}^{2},
            l_{2}^{2},
            (l_{1} - p_{1})^{2},
            (l_{1} + p_{2})^{2},
            (l_{2} + p_{3})^{2},
        $
        \newline
        $
            (l_{1} - l_{2} + p_{2} - p_{3})^{2},
            (l_{1} - l_{2} - p_{1})^{2} - m_{W}^{2},
            l_{2} \cdot p_{1},
            l_{2} \cdot p_{2}
        $
        \\
        \cline{2-3}
        & \multirow{2}{*}{3} &
        $
            l_{1}^{2},
            l_{2}^{2},
            (l_{1} + p_{3})^{2},
            (l_{1} - l_{2} + p_{3})^{2},
            (l_{1} - l_{2} - p_{2} + p_{3})^{2},
        $
        \newline
        $
            (l_{1} - p_{1} - p_{2} + p_{3})^{2} - m_{t}^{2},
            (l_{2} - p_{1})^{2} - m_{W}^{2},
            l_{2} \cdot p_{2},
            l_{2} \cdot p_{3}
        $
        \\
        \cline{2-3}
        & \multirow{2}{*}{4} &
        $
            l_{1}^{2},
            l_{2}^{2},
            (l_{1} - p_{1})^{2},
            (l_{1} + p_{2})^{2},
            (l_{1} - l_{2} - p_{1})^{2},
        $
        \newline
        $
            (l_{1} - l_{2} + p_{2} - p_{3})^{2} - m_{t}^{2},
            (l_{2} + p_{3})^{2} - m_{W}^{2},
            l_{2} \cdot p_{1},
            l_{2} \cdot p_{2}
        $
        \\
        \cline{2-3}
        & \multirow{2}{*}{5} &
        $
            l_{2}^{2},
            (l_{1} - p_{1})^{2},
            (l_{2} + p_{3})^{2},
            (l_{1} - l_{2} - p_{1})^{2},
            (l_{1} + p_{2})^{2} - m_{t}^{2},
        $
        \newline
        $
            (l_{1} - l_{2} + p_{2} - p_{3})^{2} - m_{t}^{2},
            l_{1}^{2} - m_{W}^{2},
            l_{2} \cdot p_{1},
            l_{2} \cdot p_{2}
        $
        \\
        \cline{2-3}
        & \multirow{2}{*}{6} &
        $
            l_{1}^{2},
            l_{2}^{2},
            (l_{2} - p_{1})^{2},
            (l_{1} + p_{3})^{2},
            (l_{1} - l_{2} - p_{2} + p_{3})^{2} - m_{t}^{2},
        $
        \newline
        $
            (l_{1} - p_{1} - p_{2} + p_{3})^{2} - m_{t}^{2},
            (l_{1} - l_{2} + p_{3})^{2} - m_{W}^{2},
            l_{2} \cdot p_{2},
            l_{2} \cdot p_{3}
        $
        \\
        \hline \hline
    \end{tabular}
    \caption{Definitions of the integral families.
      $l_{1}$ and $l_{2}$ are loop momenta while $p_{1}$, $p_{2}$, and $p_{3}$
      are external momenta defined in Eq.~\eqref{eq:process}.
        The remaining 9 families can be obtained by crossing $p_{1} \leftrightarrow - p_{3}$.
    }
    \label{tab:families}
\end{table}

The master integrals are defined as follows
\begin{equation}
    I(a_{1}, \ldots, a_{9})
    =
    \int \left( \prod_{n = 1}^{2} e^{\epsilon \gamma_{E}} \frac{\mathrm{d}^{d} l_{n}}{i \pi^{d / 2}} \right)
    \frac{1}{D_{1}^{a_{1}} D_{2}^{a_{2}} \cdots D_{9}^{a_{9}}}
    \text{,}
\end{equation}
where denominators $D_{i}$ can be deduced from  Table~\ref{tab:families} for each of the integral families. 
Note that we absorb a factor of $- i (4 \pi)^{2 - \epsilon} e^{\epsilon \gamma_{E}}$ per loop
into the definition of the master integrals.

\begin{figure}[ht]
    \centering
    \begin{subfigure}[ht]{0.3\linewidth}
        \centering
        \includegraphics[width=0.8\linewidth]{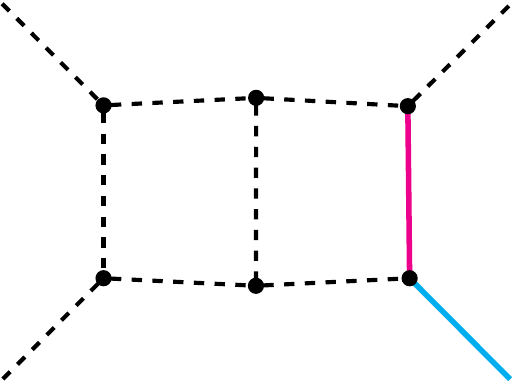}
        \caption{planar no.\ 1}
    \end{subfigure}
    ~
    \begin{subfigure}[ht]{0.3\linewidth}
        \centering
        \includegraphics[width=0.8\linewidth]{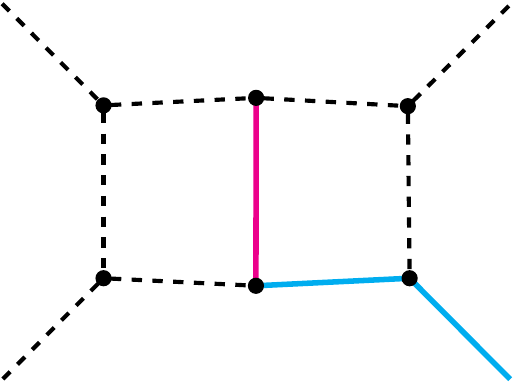}
        \caption{planar no.\ 2}
    \end{subfigure}
    ~
    \begin{subfigure}[ht]{0.3\linewidth}
        \centering
        \includegraphics[width=0.8\linewidth]{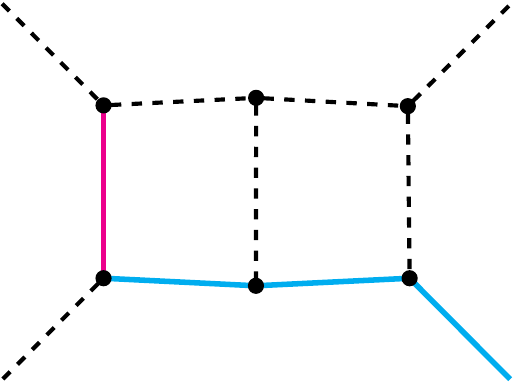}
        \caption{planar no.\ 3}
    \end{subfigure}
    \\
    \begin{subfigure}[ht]{0.3\linewidth}
        \centering
        \includegraphics[width=0.8\linewidth]{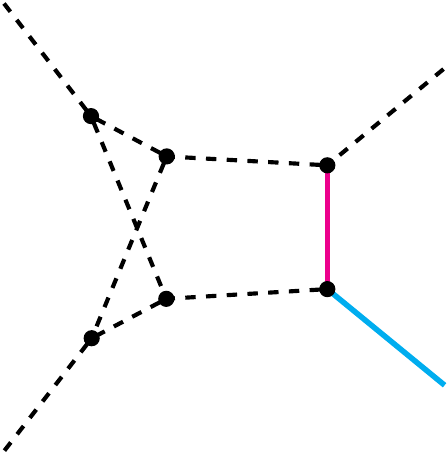}
        \caption{non-planar no.\ 1}
    \end{subfigure}
    ~
    \begin{subfigure}[ht]{0.3\linewidth}
        \centering
        \includegraphics[width=0.8\linewidth]{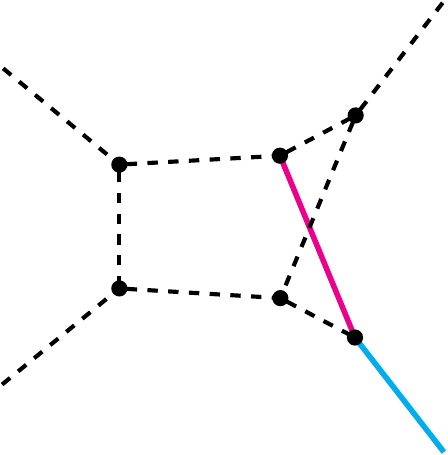}
        \caption{non-planar no.\ 2}
    \end{subfigure}
    ~
    \begin{subfigure}[ht]{0.3\linewidth}
        \centering
        \includegraphics[width=0.8\linewidth]{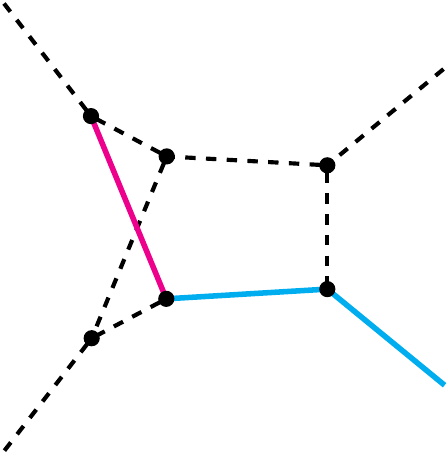}
        \caption{non-planar no.\ 3}
    \end{subfigure}
    \\
    \begin{subfigure}[ht]{0.3\linewidth}
        \centering
        \includegraphics[width=0.8\linewidth]{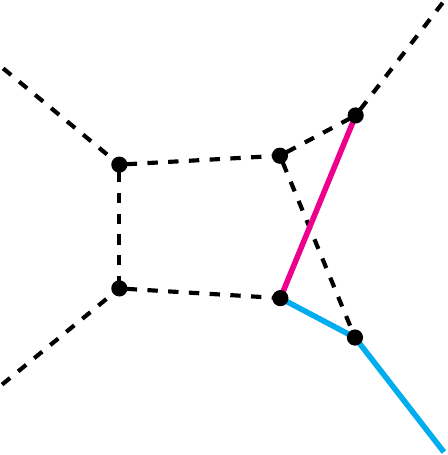}
        \caption{non-planar no.\ 4}
    \end{subfigure}
    ~
    \begin{subfigure}[ht]{0.3\linewidth}
        \centering
        \includegraphics[width=0.8\linewidth]{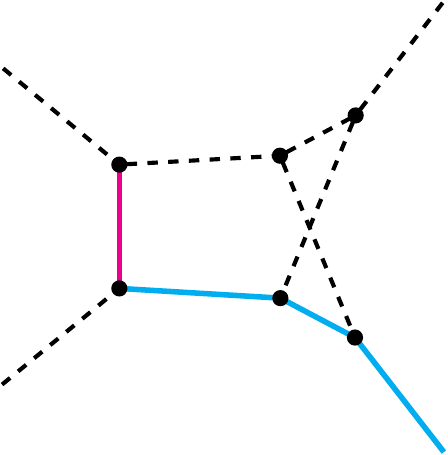}
        \caption{non-planar no.\ 5}
    \end{subfigure}
    ~
    \begin{subfigure}[ht]{0.3\linewidth}
        \centering
        \includegraphics[width=0.8\linewidth]{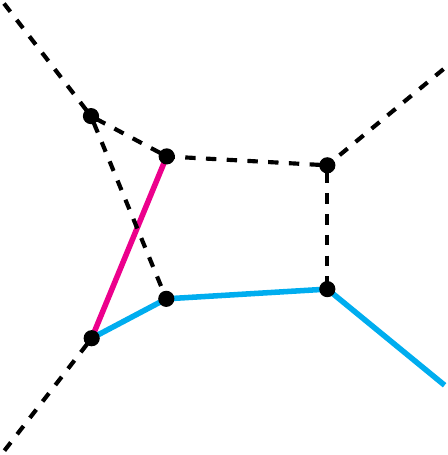}
        \caption{non-planar no.\ 6}
    \end{subfigure}
    \caption{Topologies of integral families.
        Solid and dashed lines correspond to massive and massless particles respectively.
        Blue lines have mass $m_{t}$ while red lines have mass $m_{W}$.
        All families can be crossed ($p_{1} \leftrightarrow - p_{3}$) giving a total of 18 topologies.}
    \label{fig:family}
\end{figure}

\begin{figure}[h]
    \centering
    \begin{subfigure}[ht]{0.14\linewidth}
        \centering
        \includegraphics[width=\linewidth]{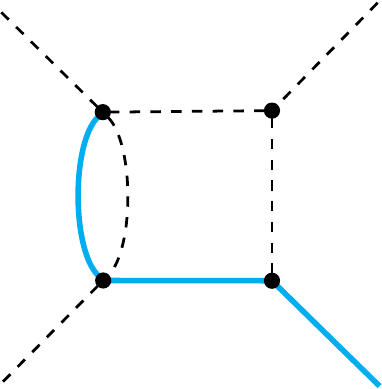}
    \end{subfigure}
    ~
    \begin{subfigure}[ht]{0.14\linewidth}
        \centering
        \includegraphics[width=\linewidth]{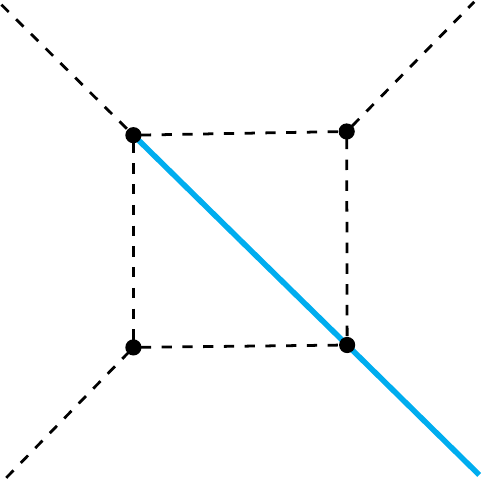}
    \end{subfigure}
    ~
    \begin{subfigure}[ht]{0.14\linewidth}
        \centering
        \includegraphics[width=\linewidth]{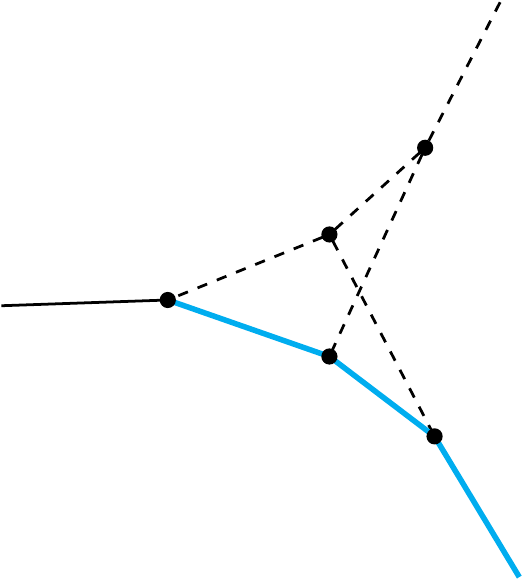}
    \end{subfigure}
    ~
    \begin{subfigure}[ht]{0.14\linewidth}
        \centering
        \includegraphics[width=\linewidth]{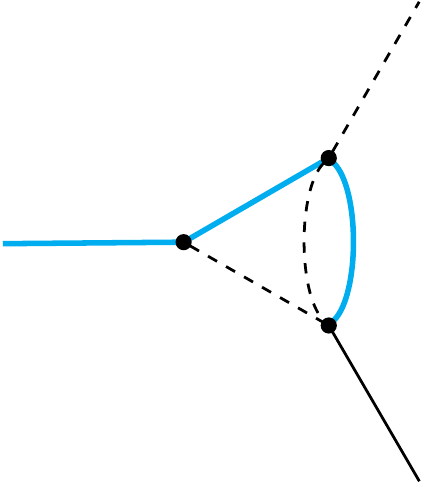}
    \end{subfigure}
    \caption{Examples of integrals that appear in the calculation of boundary conditions at $m_W^2 \to -i\infty$. 
        Solid and dashed lines correspond to massive and massless particles respectively.
        Blue lines have mass $m_{t}$ while black lines correspond to massive external particles.
    }
    \label{fig:boundaryexample}
\end{figure}

\begin{figure}[h]
    \centering
    \begin{subfigure}[ht]{0.14\linewidth}
        \centering
        \includegraphics[width=0.5\linewidth]{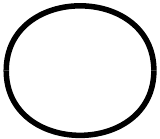}
        \caption{$I_{1}$}
    \end{subfigure}
    ~
    \begin{subfigure}[ht]{0.14\linewidth}
        \centering
        \includegraphics[width=0.5\linewidth]{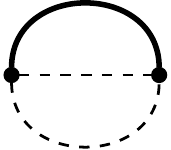}
        \caption{$I_{2}$}
    \end{subfigure}
    ~
    \begin{subfigure}[ht]{0.14\linewidth}
        \centering
        \includegraphics[width=\linewidth]{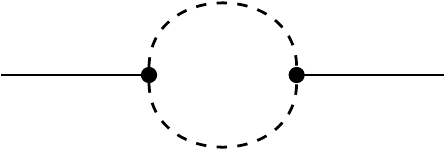}
        \caption{$I_{3}$}
    \end{subfigure}
    ~
    \begin{subfigure}[ht]{0.14\linewidth}
        \centering
        \includegraphics[width=\linewidth]{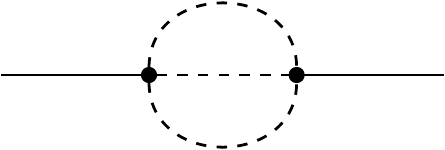}
        \caption{$I_{4}$}
    \end{subfigure}
    ~
    \begin{subfigure}[ht]{0.14\linewidth}
        \centering
        \includegraphics[width=\linewidth]{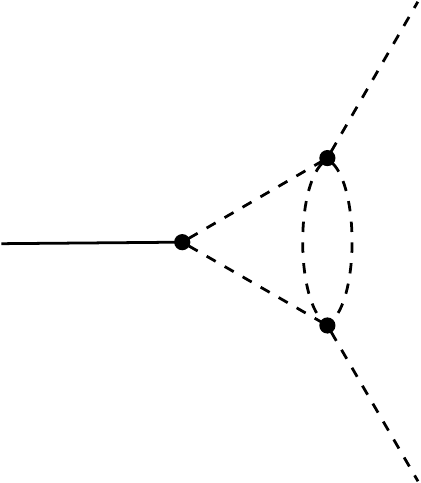}
        \caption{$I_{5}$}
    \end{subfigure}
    ~
    \begin{subfigure}[ht]{0.14\linewidth}
        \centering
        \includegraphics[width=\linewidth]{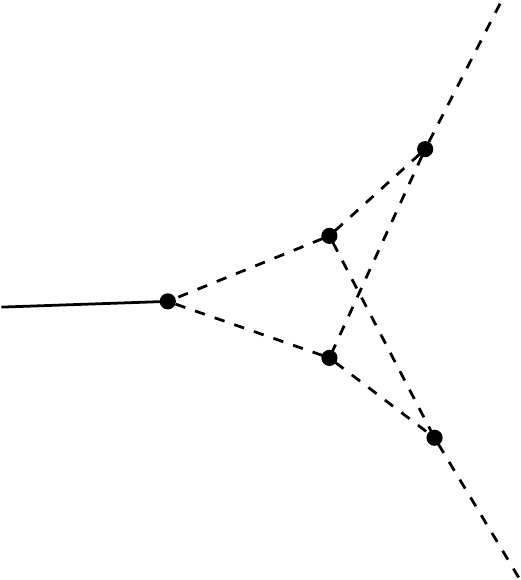}
        \caption{$I_{6}$}
    \end{subfigure}
    \\
    \begin{subfigure}[ht]{0.14\linewidth}
        \centering
        \includegraphics[width=\linewidth]{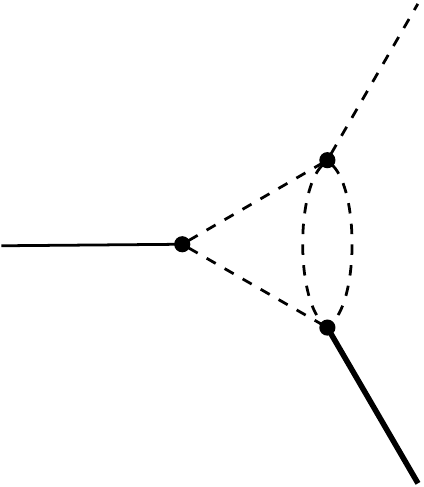}
        \caption{$I_{7}$}
    \end{subfigure}
    ~
    \begin{subfigure}[ht]{0.14\linewidth}
        \centering
        \includegraphics[width=\linewidth]{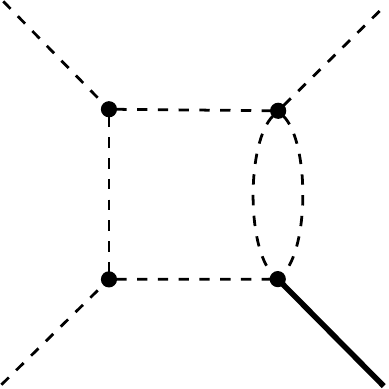}
        \caption{$I_{8}$}
    \end{subfigure}
    ~
    \begin{subfigure}[ht]{0.14\linewidth}
        \centering
        \includegraphics[width=\linewidth]{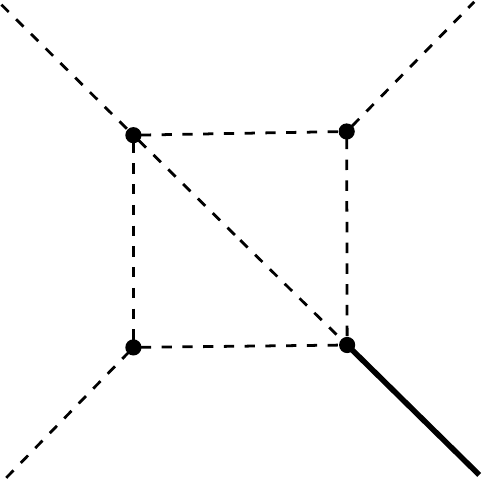}
        \caption{$I_{9}$}
    \end{subfigure}
    ~
    \begin{subfigure}[ht]{0.14\linewidth}
        \centering
        \includegraphics[width=\linewidth]{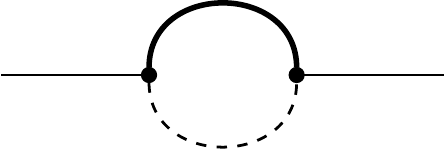}
        \caption{$I_{10}$}
    \end{subfigure}
    ~
    \begin{subfigure}[ht]{0.14\linewidth}
        \centering
        \includegraphics[width=\linewidth]{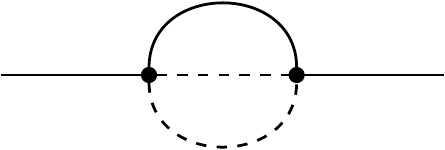}
        \caption{$I_{11}$}
    \end{subfigure}
    ~
    \begin{subfigure}[ht]{0.14\linewidth}
        \centering
        \includegraphics[width=\linewidth]{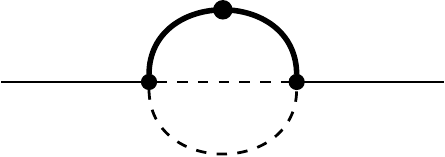}
        \caption{$I_{12}$}
    \end{subfigure}
    \\
    \begin{subfigure}[ht]{0.14\linewidth}
        \centering
        \includegraphics[width=\linewidth]{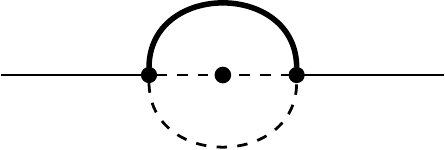}
        \caption{$I_{13}$}
    \end{subfigure}
    ~
    \begin{subfigure}[ht]{0.14\linewidth}
        \centering
        \includegraphics[width=\linewidth]{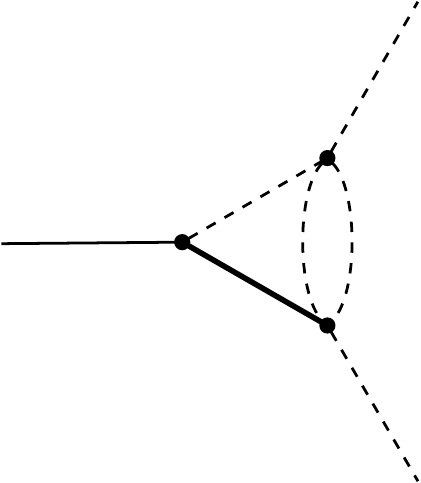}
        \caption{$I_{14}$}
    \end{subfigure}
    ~
    \begin{subfigure}[ht]{0.14\linewidth}
        \centering
        \includegraphics[width=\linewidth]{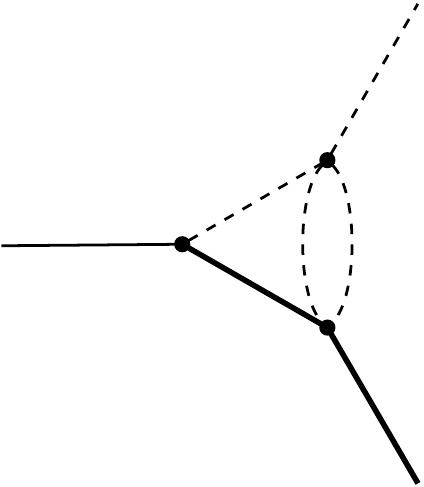}
        \caption{$I_{15}$}
    \end{subfigure}
    ~
    \begin{subfigure}[ht]{0.14\linewidth}
        \centering
        \includegraphics[width=\linewidth]{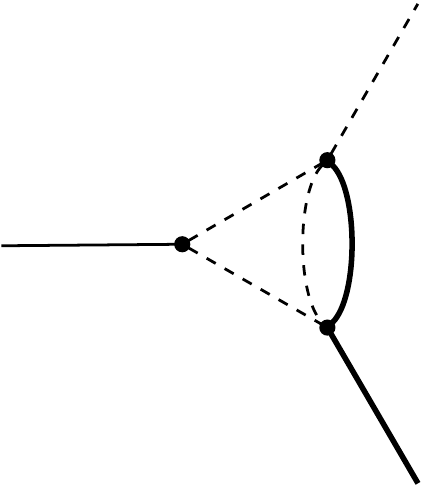}
        \caption{$I_{16}$}
    \end{subfigure}
    ~
    \begin{subfigure}[ht]{0.14\linewidth}
        \centering
        \includegraphics[width=\linewidth]{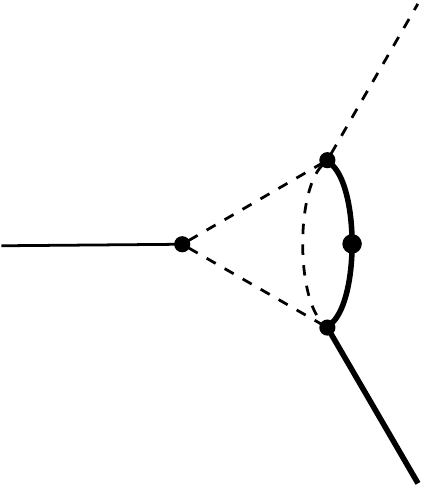}
        \caption{$I_{17}$}
    \end{subfigure}
    \caption{Master integrals for the boundary conditions.
        Solid and dashed lines correspond to massive and massless particles respectively.
        Thick solid lines represent particles with either mass $m$ or $m_t$ depending on whether the line
        is external or internal. If in some integrals thick solid lines appear both as external and internal, the mass is $m_t$. 
        Thin solid lines correspond to external particles with the momentum squared  $q^{2}$ where in general $q$ is a linear combination
    of external momenta $p_{1,2,3}$.}
    \label{fig:boundary}
\end{figure}

The calculation  of the master integrals needed to compute the non-factorisable corrections to single-top production
is complicated as they depend on Mandelstam variables and on  two masses, $m_W$ and $m_t$. We believe that, currently, 
their analytic computation is not possible. For this reason, 
we employ the auxiliary mass flow method~\cite{Liu:2017jxz,Liu:2020kpc,Liu:2021wks}
to calculate  them. To this end, 
we first construct a system of differential equations with respect to $m_{W}^{2}$,
solve it starting from the boundary conditions at $m_{W}^{2} \to -i \infty$ as required by the causality prescription,
and move to the physical value $m_{W} = 80.379 \text{~GeV}$.
To do so, we require boundary conditions at $m_{W}^{2} \to -i \infty$.
Although many integrals in this limit can be computed, we
find that some of the boundary integrals are either hard to calculate analytically 
or that analytic results available in
the literature are not known to sufficiently high orders  in the
$\epsilon$-expansion.  Examples of such integrals are shown in Figure~\ref{fig:boundaryexample}.
We take a pragmatic approach and calculate these integrals numerically.
Having already taken the limit $m_{W}^{2} \to -i \infty$, we analytically continue
$m_{t}^{2}$ in \emph{internal} propagators to the complex plane,
as the causality prescriptions  differ for internal and external masses.
We proceed as follows. 
First, we rename the top mass $m_{t}$ that appears in internal 
propagators to  $m$ and construct a system of differential equations with respect to $m^{2}$.
We solve these equations starting at the boundary  $m^{2} \to -i \infty$ and
moving  to the physical value $m = m_{t} = 173 \text{~GeV}$.  We then use these results as boundary conditions
for differential equations with respect to  $m_W^2$ at $m_W^2 = -i \infty$.
The integrals shown in Figure~\ref{fig:boundary} is the complete set that we used as boundary conditions either at $m_W^2 = -i \infty$ or $m^2 \to -i\infty$.  We note that  these
integrals can be found in an ancillary file.
In compiling this list, we have used results of Refs.~\cite{tHooft:1978jhc, Chetyrkin:1980pr, Scharf:1993ds, Gehrmann:1999as, Gehrmann:2005pd}.
We have calculated two of the master integrals ($I_{16}$ and $I_{17}$ in Figure~\ref{fig:boundary})  since we could not find them in the literature. 

For phenomenology,  we need to compute master integrals for many kinematic points relevant for the description of the process
$u + b \to d + t$.   To do that, we
can simply solve the differential equations with respect to $m_{W}^{2}$ starting at  $m_W^2 = -i\infty$
for each pair  of Mandelstam variables $s$ and $t$.
This is the approach used in the previous papers  by two of
the present authors~\cite{Bronnum-Hansen:2020mzk,Bronnum-Hansen:2021olh}.
Alternatively, we  can compute master integrals at a few kinematic points by solving differential equations
in $m_W^2$  and then use these points as boundary conditions for differential equations 
with respect to the kinematic invariants  $s$ and $t$ to calculate  master integrals at other  phase-phase points.
We note that 
a similar approach has already been used  in the literature~\cite{Maltoni:2018zvp,Frellesvig:2019byn,Abreu:2020jxa}.
In the current  calculation, we first generate several reference points in the phase space by solving the $m_{W}^{2}$ equation.
Then  we solve the equations in $s$ or $t$ to move from one of the
reference points to the point of interest.

In general there are singularities in differential  equations with respect to Mandelstam invariants; 
some of these singularities  may appear as curves in the physical phase space.
We need to use the correct causality prescription to cross such curves to 
avoid ending up on  the unphysical sheet of the Riemann surface.
One virtue of the auxiliary mass flow method is that the negative imaginary part of
the mass provides a correct way to cross   singular curves involving that mass.
Whenever we encounter such a singular curve, we can move to
the complex {\it mass}  plane using the corresponding equation,
then solve the $s$ and $t$ equations, and finally move back to the physical value of  the $W$ boson mass.
Evaluating all 428 master integrals to a precision of
twenty digits at a typical phase-space point takes less than half an hour on a single CPU core.

We perform two checks to verify the integrals computed using the method described above.
First, we  calculate the master integrals  by directly integrating over Feynman parameters
using the publicly available program 
\texttt{pySecDec}~\cite{Borowka:2017idc,Borowka:2018goh}. We perform a comparison 
at a physical phase-space point, away from kinematic thresholds\footnote{
  We  use $s = (500 \text{~GeV})^{2}$ and    $t = - (100 \text{~GeV})^{2}$.}
and   find good agreement of our  and \texttt{pySecDec} results 
for the majority of the master integrals.  Unfortunately,
for some master integrals required in this paper, in particular non-planar boxes, 
we were  unable to produce meaningful results  with \texttt{pySecDec}.

Second, we have also checked the self-consistency of the differential equations.
Indeed, solving the differential equations in $s$ and $t$ variables to move from
one phase-space point to the next should produce the same results as solving the
$m_{W}^{2}$-equations and directly moving from   $m_{W}^{2} = -i \infty$  to the phase-space point of interest.
We have checked that for several points across the phase space,
master integrals evaluated in these two different ways agree up to the target  precision
of twenty digits.

\section{Divergences in non-factorisable contributions to the scattering  amplitude}\label{sec:polestruct}

In this section we discuss the $\ep$-pole structure of non-factorisable contributions to the scattering amplitude
for the process $u+b \to d +t$. In principle, divergences of  scattering amplitudes at higher orders in perturbative
QCD are very well understood, see e.g. Refs.~\cite{Catani:1998bh,Czakon:2014oma, Becher:2009qa}.
However, since in this paper we only deal with non-factorisable
contributions to the two-loop amplitude, the relevant pole structure turns out to be quite special and, in fact, simpler
than the general case. 

Indeed, the first point to appreciate is that there are  no ultraviolet divergences  in non-factorisable corrections 
 that contribute to the  interference of the two-loop amplitude and
 the tree-level amplitude in Eq.~\eqref{eq:interference}.  Therefore, we only need to use
 a relation between the bare and
renormalised $\overline{\text{MS}}$ QCD coupling to zeroth order in perturbative QCD. It reads
\begin{align}
  \as^{\rm bare} = \mu^{2 \epsilon} S_\epsilon  \as,\label{eq:ashat}
\end{align}
where $S_\epsilon = (4\pi)^{-\epsilon} \exp (\epsilon \gamma_E)$ and $\gamma_E \approx 0.57721$
is the Euler-Mascheroni constant.  Starting with the expression for individual Feynman diagrams,
where the bare coupling constant enters,  and rewriting it using Eq.~\eqref{eq:ashat}, we obtain
the amplitude introduced in Eq.~\eqref{eq:asexp}. 

In general, after renormalisation, the only $\ep$-poles present in the amplitude
${\cal A}$ are of infrared origin.   To extract them, we follow Refs.~\cite{Catani:1998bh,Czakon:2014oma, Becher:2009qa}.
 To this end, we interpret the {\it full} amplitude
in Eq.~\eqref{eq:asexp} as a vector in colour space. This amplitude is divergent; if poles of this
amplitude are removed using the ${\overline {\rm MS}}$ prescription, we obtain a finite
amplitude $\ket{\mathcal{F}}$ that is also a vector in colour space.  Similar to the original
amplitude $\ket{\mathcal{A}}$,  $\ket{\mathcal{F}}$ can also be expanded in powers of $\alpha_s$. 
We write 
\be
\ket{\mathcal{A}} = \boldsymbol{Z} \ket{\mathcal{F}},
\ee
where $\boldsymbol{Z}$ is an operator that removes infrared poles from the amplitude $\ket{\mathcal{A}}$.
This operator satisfies  the renormalisation group equation
\begin{equation}
  \mu\frac{\,d}{\,d\mu}\boldsymbol{Z} = -\anodim \boldsymbol{Z},
  \label{eq:rge}
\end{equation}
where $\anodim$ is the so-called anomalous dimension operator.
It  reads~\cite{Aybat:2006mz,Becher:2009kw,Czakon:2009zw,Mitov:2009sv,Ferroglia:2009ii,Mitov:2010xw}
\begin{align}
   \begin{split}
        \anodim(\{p_i\},\mt,\mu) =& \sum_{(i,j)} \frac{\boldsymbol{T}_{i}\cdot
        \boldsymbol{T}_{j}}{2} \gamma_{\text{cusp}}(\alpha_s)
                                 L_{ij}
                                 + \sum_{(I,j)} \boldsymbol{T}_I\cdot
                                 \boldsymbol{T}_{j} \gamma_{\text{cusp}}(\alpha_s)
                                 L_{Ij}^{(m)} \\
                               &- \sum_{(I,J)} \frac{1}{2} \boldsymbol{T}_I\cdot
                                 \boldsymbol{T}_J \gamma_{\text{cusp}}(\nu_{IJ}, \as)
                               + \sum_{i} \gamma^{i}(\as)
                               + \sum_{I} \gamma^I\left(\as\right) \\
                               &+ \sum_{(I,J,K)} i f^{abc} T^a_I T^b_J T^c_K F_I
                               \left(\nu_{IJ},\nu_{JK},\nu_{KI}\right) \\
                               &+ \sum_{(I,J)}\sum_{k}i f^{abc}T_I^a T^b_JT^c_{k}
                     f_2\left(\nu_{IJ},\ln\left(\frac{-\sigma_{Jk}v_J \cdot p_{k}}{-\sigma_{Ik}v_I\cdot p_{k}}\right)\right)
                                  \,.
\label{eq:anodim}
\end{split}
\end{align}
Small-letter indices refer to massless external partons whereas capital-letter  indices denote massive external partons
\cite{Becher:2009kw}.
When indices in a sum are shown  in parenthesis, as e.g.  $(i,j)$, the summation should be restricted
to distinct indices.   Also, $ L_{ij} = \ln\left( \mu^2/(-s_{ij}) \right)$ and $L_{Ij}^{(m)} = \ln\left( \mt \mu /(-s_{I j}) \right)$.
Furthermore, the kinematic invariants that appear in the above equation are defined as
\begin{equation}
    s_{ij} = 2\sigma_{ij} p_i\cdot p_j + i \varepsilon
    \,,
\end{equation}
where $\sigma_{ij} = 1$ if both $p_i$ and $p_j$ are incoming or outgoing and $\sigma_{ij}=-1$ otherwise.
Quantities $\nu_{IJ}$ are cusp angles, $v_I$ are four-velocities and $ \boldsymbol{T}_i$ are   colour-charge
operators of the corresponding partons. 

The operator  $\anodim$  in Eq.~\eqref{eq:anodim} describes infrared and collinear divergences of the full 
amplitude that includes both factorisable and non-factorisable terms. However, if we focus on non-factorisable
contributions only, the expression for $\anodim$  can be simplified. 
Indeed, we note that the last two terms in Eq.~\eqref{eq:anodim} are not needed for predicting
infrared poles  of  the non-factorisable amplitude $\mathcal{A}_{\rm nf}^{(2)}$ 
since they are proportional to non-abelian colour factors.  As we have explained in Section~\ref{sec:ampcalc}, non-abelian
colour factors cannot arise in the contributions that we are interested in. 
Furthermore, the two sums over the anomalous dimensions $\gamma^{(i)}$ and $\gamma^{(I)}$
cannot  contribute to $\mathcal{A}_{\rm nf}^{(2)}$ either since they are related to collinear emissions
that, in physical gauges, should   be associated with   factorisable parts of  the amplitude.

We are left  with three sums that involve  products of two colour-charge operators in Eq.~\eqref{eq:anodim}.
In our case, there are three  massless and one massive external particle; hence,  the third sum
in Eq.~\eqref{eq:anodim} that should be performed
over two distinct massive indices is not relevant for us and can be discarded. 
Moreover, the non-factorisable contributions involve gluon exchanges between different fermion lines. 
Hence only  four  products of colour-charge operators
contribute; they are   $\boldsymbol{T}_1 \cdot \boldsymbol{T}_2$,
$\boldsymbol{T}_1 \cdot \boldsymbol{T}_4$,
$\boldsymbol{T}_2 \cdot \boldsymbol{T}_3$, and $\boldsymbol{T}_3 \cdot \boldsymbol{T}_4$.
Finally, the cusp  anomalous dimension is given by~\cite{Becher:2009kw,Becher:2009cu}
\begin{equation}
        \gamma_{\text{cusp}} = 4 \left(\frac{\as}{4\pi}\right) + \left[ \left(\frac{268}{9}-\frac{4\pi^2}{3}\right)C_A - \frac{80}{9}T_F n_l\right] \left(\frac{\as}{4\pi}\right)^2 + \mathcal{O}(\as^3)
        \,.
\end{equation}
We note that the ${\cal O}(\alpha_s^2)$  contribution to $\gamma_{\text{cusp}}$
contains terms that are either  proportional to a non-abelian colour factor $C_A$ or to the number
of light fermions $n_f$  and none of these parameters appear in the non-factorisable diagrams.
Hence, if we are interested in non-factorisable corrections only, the
$C_A$- and $n_f$-dependent  contributions to $\gamma_{\text{cusp}}$ should be discarded. Therefore,
we are allowed to replace
\begin{equation}
\gamma_{\text{cusp}} \to  \gamma_{\rm nf} = 4 \left(\frac{\as}{4\pi}\right),
\end{equation}
in the expression for $\anodim$ in Eq.~\eqref{eq:anodim}.

We define the part of the operator $\anodim$ that is relevant for non-factorisable corrections as  $\anodim_{\rm nf}$.
As a consequence of the above discussion, it  reads 
\begin{align}
  \anodim_{\rm nf}(\{p_i\},\mt,\mu) &= \left(\frac{\as}{4\pi}\right) \anodim_{0,\rm nf}(\{p_i\},\mt,\mu),
\end{align}
where 
\begin{align}
\anodim_{0,\rm nf} & = 
    4 \bigg[
      \boldsymbol{T}_{1}\cdot\boldsymbol{T}_{2}\ln\left(\frac{\mu^2}{-s-i\varep}\right)
      + \boldsymbol{T}_{2}\cdot\boldsymbol{T}_{3}\ln\left(\frac{\mu^2}{-u-i\varep}\right)\\
     &\quad+ \boldsymbol{T}_{1}\cdot\boldsymbol{T}_{4}\ln\left(\frac{\mu\:\mt}{\mt^2 - u -i\varep}\right)
      + \boldsymbol{T}_{3}\cdot\boldsymbol{T}_{4}\ln\left(\frac{\mu\:\mt}{\mt^2-s-i\varep}\right)
    \bigg].
\end{align}

We can solve Eq.~\eqref{eq:rge} with $\anodim_{\rm nf}$ in place of $\anodim$ order by order in $\alpha_s$, 
to determine the operator $ \boldsymbol{Z}_{\rm nf}$; we assume that  $\boldsymbol{Z}_{\rm nf}  = 1 + {\cal O}(\alpha_s)$.
This solution is much simpler than the one  for the full amplitude since perturbative  running of
the coupling constant cannot play a role in the non-factorisable contribution through ${\cal O}(\alpha_s^2)$.
As the result, we find a remarkably simple expression 
\begin{align}
  \boldsymbol{Z}_{\rm nf} = 1 + \left ( \frac{\alpha_s}{4\pi} \right ) \frac{\anodim_{0,\rm nf}}{2 \epsilon} 
  + \left (\frac{\alpha_s}{4\pi} \right )^2  \frac{{\anodim_{0,\rm nf}}^2}{8 \epsilon^2} + {\cal O}(\alpha_s^3),
  \label{eq:anodimnf}
\end{align}
which emphasizes that through two loops  non-factorisable contributions  are abelian even if computed in a non-abelian theory like QCD. 
We also note that 
in comparison to infrared  $\ep$-poles  in the full amplitude, divergences in non-factorisable contributions
are much more mild and start at $1/\ep$ at ${\cal O}(\alpha_s)$ and at $1/\ep^2$ at ${\cal O}(\alpha_s^2)$.  This is a direct
consequence of the fact that collinear divergences cannot appear in a non-factorisable amplitude due to its very definition.
Hence, all infrared poles present in Eq.~\eqref{eq:anodimnf} are of soft origin.

It is now  straightforward to predict  $\ep$-poles  in  non-factorisable contributions to cross sections. We find  
\begin{align}
\begin{split}
  \braket{\mathcal{A}^{(0)}}{\mathcal{A}_{\rm nf}^{(2)}} &= -\frac{1}{8 \ep^2} \bra{\mathcal{A}^{(0)}} \anodim_{0,\rm nf}^2 \ket{\mathcal{A}^{(0)}}
  + \frac{1}{2 \ep} \bra{\mathcal{A}^{(0)}} \anodim_{0,\rm nf} \ket{\mathcal{A}_{\rm nf}^{(1)}} + \braket{\mathcal{A}^{(0)}}{\mathcal{F}_{\rm nf}^{(2)}}, \\
  \braket{\mathcal{A}_{\rm nf}^{(1)}}{\mathcal{A}_{\rm nf}^{(1)}} &= \frac{1}{4 \ep^2} \bra{\mathcal{A}^{(0)}} \vert \anodim_{0,\rm nf} \vert^2 \ket{\mathcal{A}^{(0)}}
  + \frac{1}{2 \ep} \bra{\mathcal{A}_{\rm nf}^{(1)}} \anodim_{0,\rm nf} \ket{\mathcal{A}^{(0)}}
 \\
 &\;\;\;\;\;\;\;\;\;\;\;\;\;\;\;\;\;\;\;\;\;\;\;\;\;\;\;\;\;
 + \frac{1}{2 \ep} \bra{\mathcal{A}^{(0)}} \anodim_{0,\rm nf}^\dagger \ket{\mathcal{A}_{\rm nf}^{(1)}} + \braket{\mathcal{F}_{\rm nf}^{(1)}}{\mathcal{F}_{\rm nf}^{(1)}}.
  \label{eq:polestructure}
\end{split}
\end{align}

We can easily calculate  matrix elements of the relevant colour-charge  operators. As an example, consider
$\bra{\mathcal{A}^{(0)}} \anodim_{0,\rm nf}^2 \ket{\mathcal{A}^{(0)}}$. The action of colour-charge operators 
on vectors in the colour space is defined as follows 
\begin{align}
\bra{\boldsymbol{e}} \boldsymbol{T}_i^a \ket{\boldsymbol{d}} = T_{e_i d_i}^a \prod_{j \neq i} \delta_{e_j d_j},\quad T_{e_i d_i}^a =\begin{cases} 
t_{e_i d_i}^a & \text{final state quark}, \\
-t_{d_i e_i}^a &\text{initial state quark}.
\end{cases}
\,.
\end{align}
In the above equation
$\ket{\boldsymbol{d}}$ and $\ket{\boldsymbol{e}}$ are vectors in colour space and the  $SU(3)$ generators $t^{a,b}$ are
normalised in a standard way
\begin{equation}
        \text{Tr}\left(t^a t^b\right) = \frac{1}{2}\delta_{ab}
        \,.
\end{equation}
Using these definitions, it is easy to  see that for all combinations of colour-charge operators  that appear in $\anodim_{0,\rm nf}^2$  the
following results holds
\begin{equation}
\bra{\mathcal{A}^{(0)}}  ( \boldsymbol{T}_i \boldsymbol{T}_j )  ( \boldsymbol{T}_k \boldsymbol{T}_m ) \ket{\mathcal{A}^{(0)}}
 = (-1)^{n_i}  \frac{N_c^2-1}{4}\,.
\end{equation}
Here  $n_{i}$ is the number of indices among $i,j,k,m$
that correspond to initial-state partons. Hence, we  find 
\begin{equation}
\bra{\mathcal{A}^{(0)}} \anodim_{0,\rm nf}^2 \ket{\mathcal{A}^{(0)}}
= 4 ( N_c^2-1)  |A^{(0)}|^2
\left (
\ln\left(\frac{-u-i\varep }{-s-i\varep}\right) +
 \ln\left(\frac{m_t^2 - u -i\varep}{\mt^2-s-i\varep}\right )
   \right )^2.
\end{equation}
  All other contributions  that appear in Eq.~\eqref{eq:polestructure} can be computed in a similar way.

  Predictions for  infrared poles of the two-loop non-factorisable contribution to the cross section  provide an important
  cross check of the correctness of the calculation.  As an example of the level of numerical precision  that we have achieved for the
  $\ep$-poles of the two-loop non-factorisable amplitude, 
  in Table~\ref{tab:ampevaluation} we compare the results
  of the evaluation of $\braket{\mathcal{A}^{(0)}}{\mathcal{A}_{\rm nf}^{(2)}}$ with the  analytic predictions for its $\ep$-poles.  
   We observe that analytic and numerical  results for $\ep$-poles agree to 15 digits.
 We also find that the  $\ep$-poles of   $\braket{\mathcal{A}_{\rm nf}^{(1)}}{\mathcal{A}_{\rm nf}^{(1)}}$ are  accurate to about  30 digits
 throughout the phase space since  the one-loop integrals are evaluated to that  precision.
 In Appendix~\ref{app:evaluations} we provide additional numerical results for non-factorizable contributions,  including their
 finite parts,  for further reference.

\begin{table}[t]
	\begin{center}
		\begin{tabular}{|c|c|c|}
			\hline
			& $\epsilon^{-2}$ & $\epsilon^{-1}$ \\
			\hline \hline
			$\braket{\mathcal{A}^{(0)}}{\mathcal{A}_{\rm nf}^{(2)}}$ & \scriptsize{$-229.0940408654660 -8.978163333241640 i$} & \scriptsize{$-301.1802988944764 -264.1773596529505 i$}  \\
			IR poles & \scriptsize{$-229.0940408654665 -8.978163333241973 i$} & \scriptsize{$-301.1802988944791 -264.1773596529535 i$} \\
			\hline
		\end{tabular}
	\end{center}
	\caption{Computed and predicted $\ep$-poles 
          for a typical phase space point. We use  $s \approx 104337~\text{GeV}^2$ and $t \approx -5179.68~\text{GeV}^2$ for a comparison.}
	\label{tab:ampevaluation}
\end{table}

\section{Results}\label{sec:results}

Having computed the two-loop non-factorisable  contribution to the scattering amplitude for
single-top production, we can  study its impact on  the  single-top production cross section.
Such an analysis is  necessarily incomplete. Indeed, since in  this paper we restrict ourselves
to virtual corrections, we will have to consider quantities that depend on how the infrared
singularities are removed. To arrive at the physical result which is independent of the infrared regulator,
we need to combine virtual corrections computed in this paper with real-emission non-factorisable contributions.  
We intend to do this  in the future.   However, we believe it is still useful  to study the contribution of virtual
corrections computed in this paper to the single-top production cross section. Indeed,  as we explained
in the previous sections,  we computed  master integrals numerically.
Hence, it is important to show that
our numerical evaluation is sufficiently fast and robust to enable realistic phenomenological studies. 

To address this point, we study   non-factorisable corrections
to the differential cross section for single-top production at the LHC in the $ub$-channel. We write 
\begin{align}
    \,d\sigma^{ub}_{pp\rightarrow d + t} = \sum_{\substack{i,j = u,b \\ i \neq j}}\int\,dx_1\,dx_2\, f_i(x_1)f_j(x_2)\,d\hat{\sigma}_{ij\rightarrow d + t}(x_1, x_2) \,,
\end{align}
where $f_i$ are parton distribution functions (PDFs) and the superscript indicates that we only
consider the $ub$ initial state. 
We consider proton-proton collision at 13 TeV and
use the \texttt{NNPDF31\_lo\_as\_0118} parton distribution functions~\cite{Buckley:2014ana,NNPDF:2017mvq}.
The renormalisation and factorisation scales are fixed at $\mu=\mt$.
The value of the strong coupling constant $\as$ is provided by the PDF sets. 
  We use $m_t=173~{\rm GeV}$, $m_W = 80.379~{\rm GeV}$,
  the Fermi constant $G_F = 1.16637 \times 10^{-5}~~{\rm GeV}^{-2}$  and set CKM matrix elements to one.
  Finally, we  note that no kinematic cuts are applied. 

We compute the partonic cross section using the finite amplitude ${\mathcal{F}}$. We write 
\begin{align}
\hat{\sigma}_{ij\rightarrow d + t} = \frac{1}{8 N_c^2 \:s}\int\frac{\,d^3 p_3}{(2\pi)^3\: 2 E_3}\frac{\,d^3 p_4}{(2\pi)^3\: 2 E_4}
    {\braket{\mathcal{F}}{\mathcal{F}}}\: (2\pi)^4\delta^{(4)}\left(p_1+ p_2-p_3-p_4\right)
    \,,
\end{align}
where the prefactor on the right-hand side includes spin- and colour-averaging factors.
Since we are interested in the non-factorisable two-loop QCD contribution we use 
\begin{align}
    \braket{\mathcal{F}}{\mathcal{F}} = \braket{\mathcal{F}^{(0)}}{\mathcal{F}^{(0)}}
    + \left(\frac{\as}{4\pi}\right)^2
    \left[\braket{\mathcal{F}_{\rm nf}^{(1)}}{\mathcal{F}_{\rm nf}^{(1)}}
        + 2\Re\left\{\braket{\mathcal{F}^{(0)}}{\mathcal{F}_{\rm nf}^{(2)}}\right\} \right]
        \,.\label{eq:xsec}
\end{align}
As we already mentioned,   ${\cal O}(\as)$  non-factorisable contribution vanishes due to colour conservation.

In practice, the evaluation of the  non-factorisable contribution to the cross section proceeds as follows.
As a first step we produce  a reliable grid for the evaluation of the leading order cross section as well
as top rapidity and $p_\perp$ distributions.
Once the grid is obtained, we  randomly draw kinematic points from it, compute
$\braket{\mathcal{F}}{\mathcal{F}}$ and the phase-space weight for these points and
obtain an estimate of  the cross section including non-factorisable corrections. 

We find the following result
\be
\sigma^{ub}_{pp\to dt} = \left ( 90.3 + 0.3 \left ( \frac{\alpha_s(\mu_{\rm nf})}{0.108}  \right )^2  \right ) ~{\rm pb},
\label{eq:xsvalue}
\ee
where the first term  is the leading order cross section\footnote{The leading order cross section
has been checked against \texttt{MadGraph5\_{}aMC{@}NLO}~\cite{Alwall:2014hca}. }
and the second term is the
non-factorisable NNLO contribution.   We have indicated  in Eq.~\eqref{eq:xsvalue}  that one can change the scale
of the strong coupling constant in non-factorisable corrections independently of the rest of the calculation. 
This is so because the non-factorisable corrections appear for the first time at NNLO so that they cannot
compensate the scale variations of leading and
next-to-leading order cross sections.  This remark is important as the choice of $\mu_{\rm nf}$ in Eq.~\eqref{eq:xsvalue} has obvious consequences
for how large these corrections actually are.  
We note that the non-factorisable correction 
$0.3~{\rm pb}$  in Eq.~\eqref{eq:xsvalue}  is  the result of the cancellation between
the one-loop squared contribution ($0.7$~{\rm pb}) and the interference of the two-loop amplitude with the
leading order one ($-0.4$~{\rm pb}).

It follows from Eq.~\eqref{eq:xsvalue} that non-factorisable corrections  are  quite small;  they  change the leading order cross section
by $0.3$ percent.  However, in spite of being small they are  actually
of the same order as the \emph{factorisable} corrections to single-top production. Indeed, factorisable  corrections
are supposed to be the dominant ones but they change  the NLO single-top
production cross section by  less than a percent (see e.g. Ref.~\cite{Campbell:2020fhf}).
Moreover, as we already mentioned, the appropriate choice of the scale $\mu_{\rm nf}$ in Eq.~\eqref{eq:xsvalue} is
unclear at present.  However, since these corrections always involve exchanges between two quark lines,
it is reasonable to assume that  proper  $\mu_{\rm nf}$ should be  related to a  typical transverse momentum
 of the top quark in single-top production,  which is about $40-60~{\rm GeV}$.  If so, the magnitude of
the non-factorisable correction will increase by a factor ${\cal O}(1.5)$ and become close to half a percent.

\begin{figure}[t]
    \centering
    \includegraphics[width=0.85\linewidth]{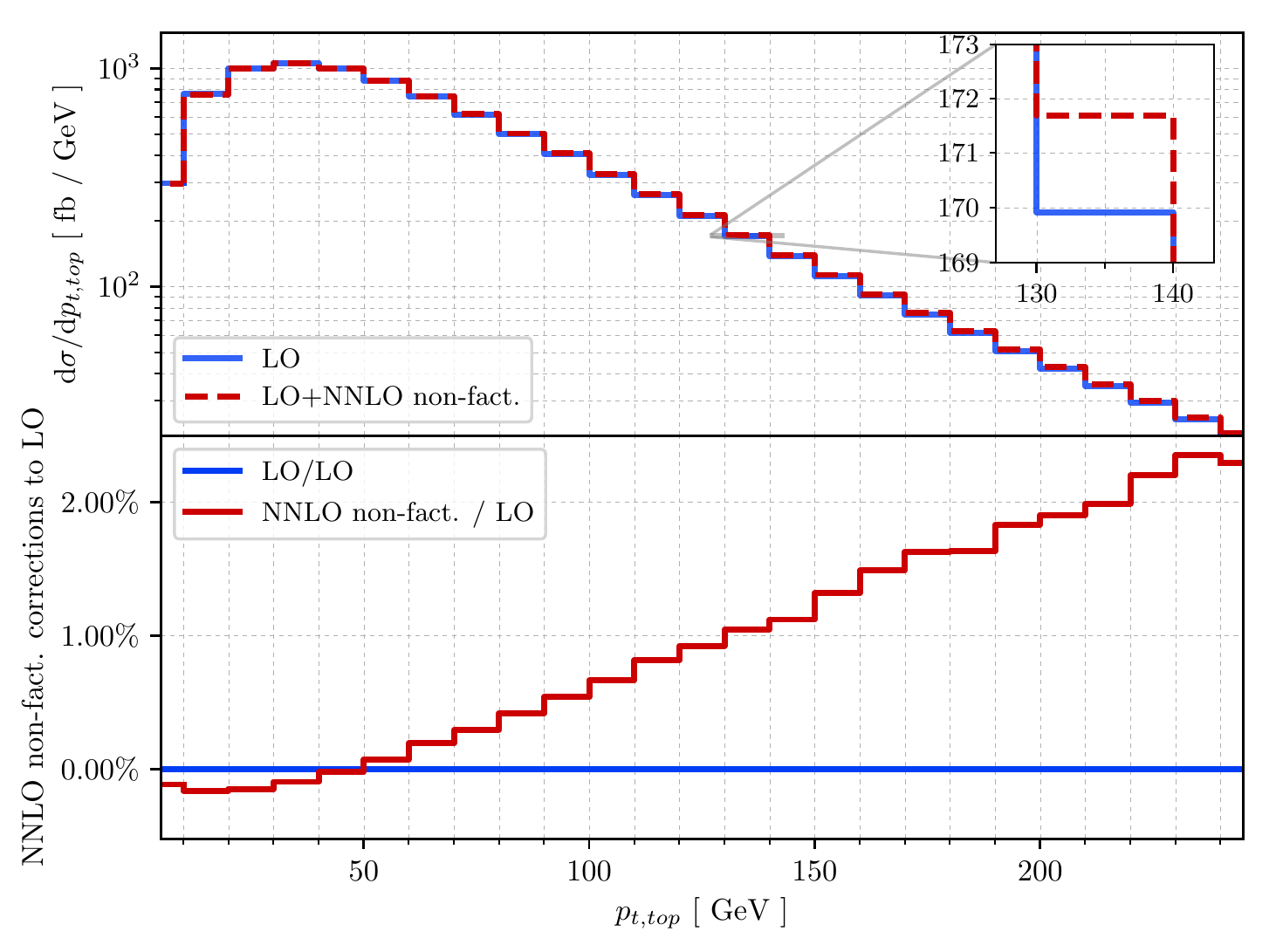}
    \caption{ The top-quark transverse momentum distribution.  In the upper pane, the blue line corresponds to  the leading order
      distribution whereas the dashed, red line to  the distribution with NNLO QCD non-factorisable corrections included. 
      In the lower pane, the ratio of non-factorisable corrections to the  leading order distribution is presented.
    See text for further details.}
    \label{fig:pttop}
\end{figure}

Having discussed the total cross section we move to   kinematic distributions. 
We begin with the distribution of  the top quark transverse momentum;  it is shown in Figure~\ref{fig:pttop}.
In the upper pane we display the differential cross section at leading order and including non-factorisable corrections.
In the lower pane, we show  ratios of the  NNLO non-factorisable correction to the leading order differential cross section
as a function of the top quark transverse momentum.

It follows from   Figure~\ref{fig:pttop} that non-factorisable corrections exhibit significant $p_\perp$-dependence.
Indeed, they are quite small and negative for $p_\perp$ between $0$ and  $50$ GeV. For larger $p_\perp$,
they start growing and reach ${\cal O}(1 \%)$ at $p_\perp \sim 100~{\rm GeV}$. 
It is interesting to note that the NNLO \emph{factorisable} correction exhibits a
similar $p_\perp$-dependence~\cite{Brucherseifer:2014ama,Berger:2016oht,Campbell:2020fhf}
which means that the relative importance
of factorisable and non-factorisable corrections remains constant across the phase space.

\begin{figure}[t]
    \centering
      \begin{subfigure}{0.495\textwidth}
    \centering
    \includegraphics[width=\textwidth]{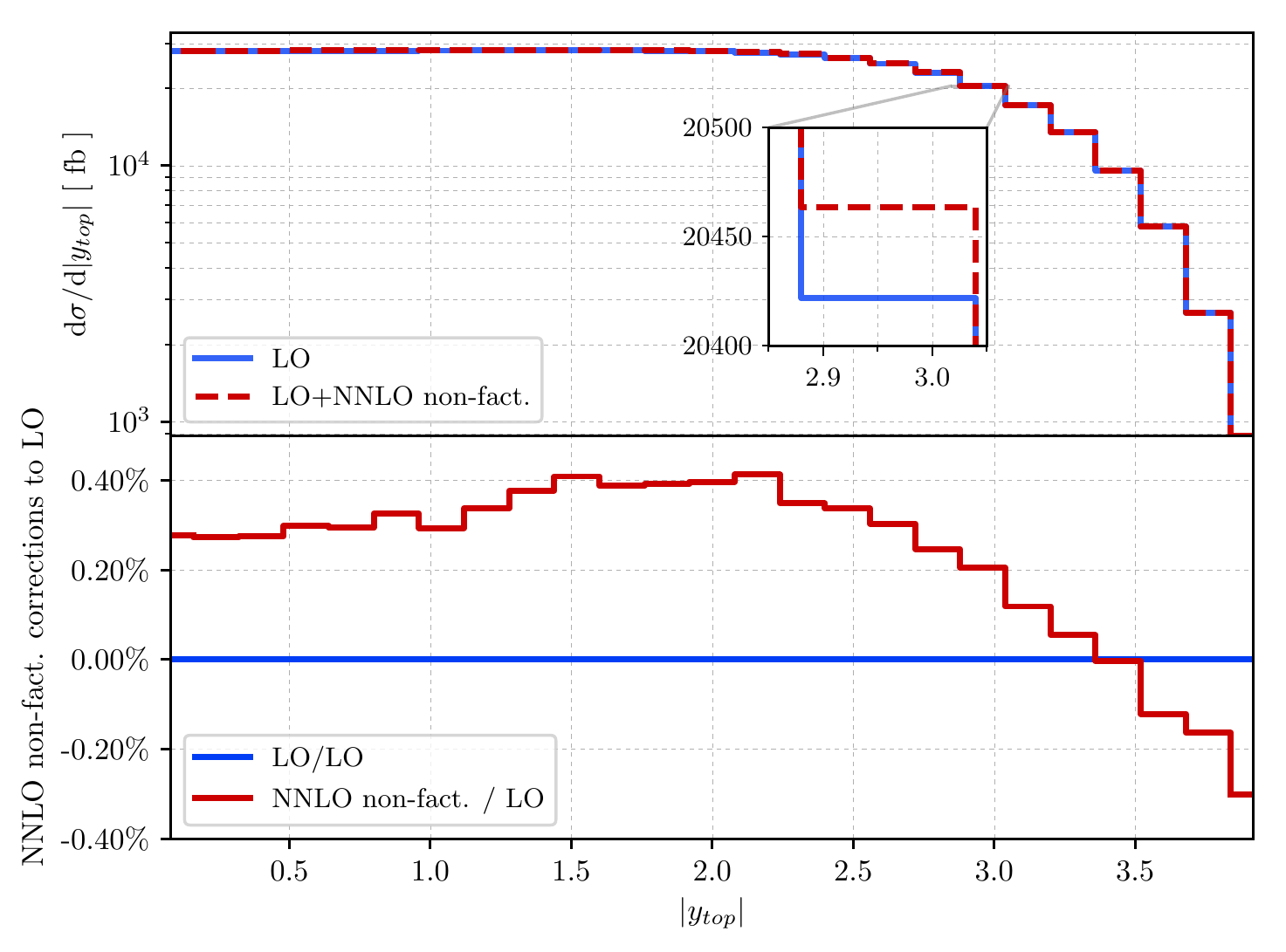}
    \caption{The top-quark rapidity distribution.}
    \label{fig:ytop}
  \end{subfigure}
  \begin{subfigure}{0.495\textwidth}
    \centering
    \includegraphics[width=\textwidth]{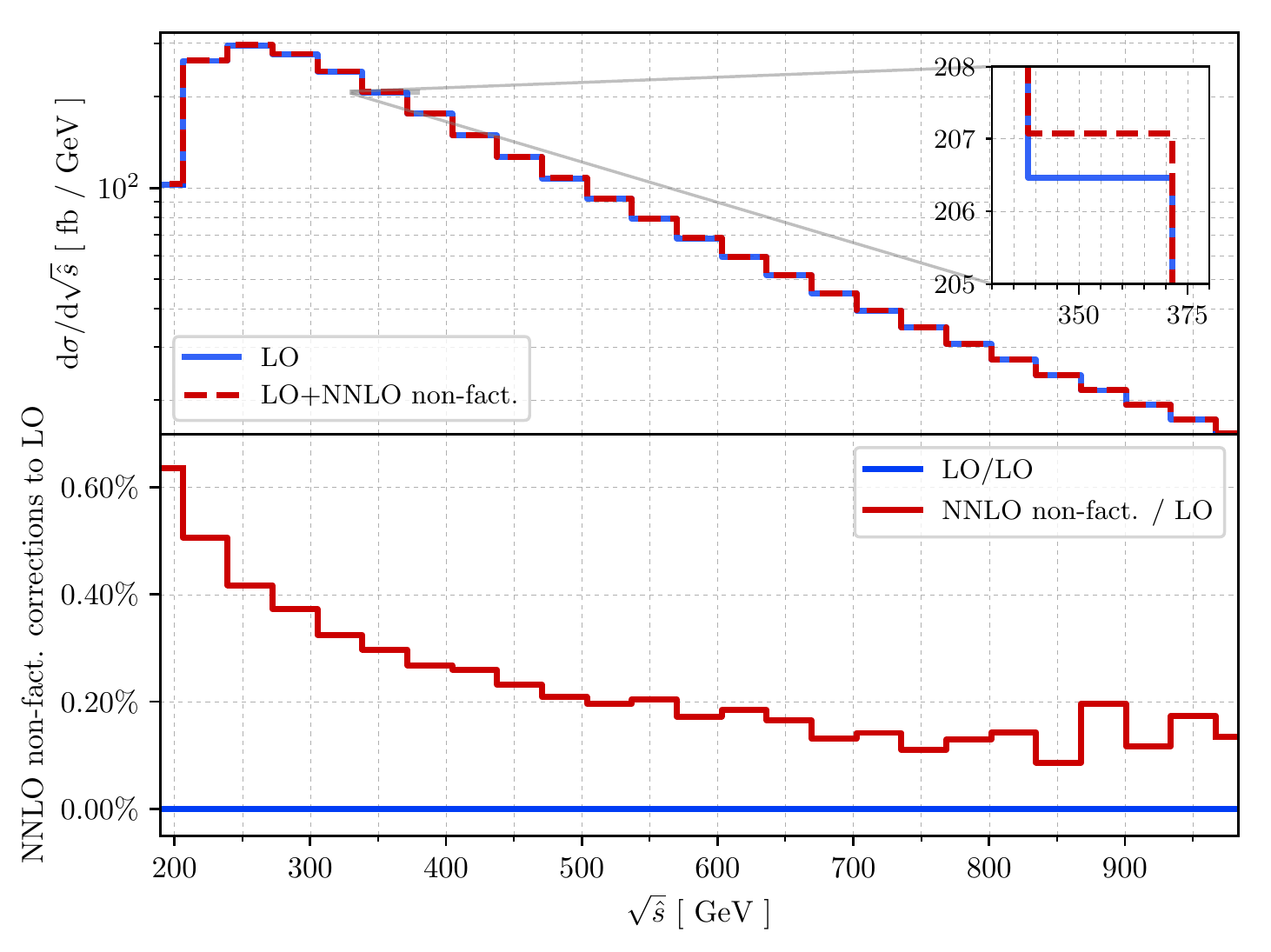}
    \caption{Differential cross section with respect to the partonic invariant energy $\sqrt{\hat{s}}$.}
    \label{fig:shat}
  \end{subfigure}
  \caption{Distributions of the  absolute value of the top-quark rapidity  (left) and the partonic center of mass
    energy $\sqrt{\hat{s}}$ (right). Upper panes show leading order distributions as well as distributions with  non-factorisable
    corrections included. 
    Lower panes show  the ratio of non-factorisable corrections to leading order distributions. See text for further details.}
  \label{fig:ytopandshat}
\end{figure}

In Figure~\ref{fig:ytopandshat}, we show  the impact of non-factorisable corrections on the top-quark rapidity distribution and
on the distribution  of the invariant mass of the top quark and the light-quark jet which for a $2 \to 2$ process is
equivalent to the partonic centre-of-mass energy  $\sqrt{\hat{s}}$.
It follows from Figure~\ref{fig:ytop} that  non-factorisable corrections to the rapidity distributions are ${\cal O}(0.3\%)$
for $|y_t| < 2.5$; for larger rapidities corrections quickly  become negative.   The non-factorisable corrections to the
$\sqrt{\hat{s}}$ distribution shown in Figure~\ref{fig:shat} 
are positive and change from ${\cal O}(0.6\%)$ at the threshold to ${\cal O}(0.1\%)$ at large partonic centre-of-mass energies.

\section{Conclusions}\label{sec:conclusions}

In this paper, we computed the contribution of two-loop non-factorisable virtual corrections 
to  $t$-channel single-top production cross section.
This is the last missing part of the two-loop amplitude  needed for a complete description of this process
through NNLO QCD.
Exact dependence on the top quark mass is retained throughout the calculation.

The calculation reported in this paper involves numerical computation of master integrals using the auxiliary mass
flow method  \cite{Liu:2017jxz,Liu:2020kpc,Liu:2021wks}. For  this reason it is important to demonstrate that
the calculation is sufficiently robust and can be used to produce results relevant for phenomenology.
We have shown this  by studying the impact of finite remainders of non-factorisable virtual corrections on the single-top production
cross section and basic kinematic distributions. We have found that non-factorisable corrections are smaller than, but quite
comparable to,  the factorisable ones,  especially since it is not very clear which scale for the strong coupling
constant should be used when computing them. 

We emphasize that  phenomenological studies reported in Section~\ref{sec:results} are necessarily incomplete since
they include only virtual non-factorisable corrections. As we explained in Section~\ref{sec:polestruct}, these virtual corrections 
are infrared divergent; hence,
to arrive at physical predictions, we also require non-factorisable real-emission contributions.
Given the recent  progress with understanding of how fully-differential NNLO QCD computations can be performed, 
we believe that it is  straightforward 
to compute the non-factorisable real-emission corrections to single-top production; we plan to
do this  in the near future.  Finally, realistic phenomenological studies require  an inclusion  of top quark decays.
Since we computed  non-factorisable contributions to two-loop helicity amplitudes, it is straightforward to accommodate top quark decays
into our computation as well.

\acknowledgments

This research is partially supported by the Deutsche Forschungsgemeinschaft (DFG, German Research Foundation) under grant 396021762 - TRR 257.
\texttt{JaxoDraw}~\cite{Binosi:2008ig} was used for the Feynman diagrams in Figure~\ref{fig:diagrams},~\ref{fig:factorisable}, and~\ref{fig:treelevel}.
The diagrams in Figure~\ref{fig:family},~\ref{fig:boundaryexample}, and~\ref{fig:boundary} were generated using \texttt{tikz-feynman}~\cite{Ellis:2016jkw}.

\appendix

\section{Numerical evaluations}\label{app:evaluations}

In Table~\ref{tab:refevals} we provide numerical results for the non-factorisable contribution to the two-loop amplitude for three kinematic points.

\begin{table}[ht]
	\begin{center}
		\begin{tabular}{|c|c|c|c|}
			\hline
			$\braket{\mathcal{A}^{(0)}}{\mathcal{A}_{\rm nf}^{(2)}}(s,t)$ & $\epsilon^{-2}$ & $\epsilon^{-1}$ & $\epsilon^{0}$ \\
			\hline \hline
			 \scriptsize{$(104337.30, -5179.6797)$} & \scriptsize{$-229.09404 -8.9781633 i$} & \scriptsize{$-301.18030 -264.17736 i$} & \scriptsize{$380.61217 + 307.59053 i$} \\
			\hline
			 \scriptsize{$(51824.679, -16060.887)$} & \scriptsize{$-8.2985887 -4.8234599 i$} & \scriptsize{$-7.2779624 -22.421862 i$} & \scriptsize{$42.503179 + 59.484685 i$} \\
			\hline
			 \scriptsize{$(2728123.9, -69809.245)$} & \scriptsize{$-5061.2720-83.997993 i$} & \scriptsize{$34392.588 -1255.7061 i$} & \scriptsize{$-1507.7598 + 18782.966 i$} \\
			\hline
		\end{tabular}
	\end{center}
	\caption{Numerical results for non-factorisable corrections at three different kinematic points specified by $(s,t)$ in units of $\text{GeV}^2$.
          For presentation purposes we have truncated numerical values to  eight digits. We use
       $m_W = 80.379~{\rm GeV}$, $m_t = 173~{\rm GeV}$ and $\mu = m_t$. See main text for further details.}\label{tab:refevals}
\end{table}

\bibliographystyle{JHEP}
\bibliography{references}

\end{document}